\documentclass{aa}
\usepackage[varg]{txfonts}
\usepackage{hyperref}
\usepackage{longtable}
\usepackage{color}
\usepackage{tikz}
\usepackage{pgfplots, pgfplotstable}
\usepackage{footnote}
\usepackage{multirow}
\usepackage[shortcuts]{extdash}
\usepgfplotslibrary{statistics}
\usetikzlibrary{patterns}
\pgfplotsset{compat=newest}
\usepackage[scientific-notation=true]{siunitx}

\begin{document}

\title{Empirical multi-wavelength prediction method for very high energy gamma-ray emitting BL Lacs\thanks{Complete version of Tables \ref{tab1} and \ref{tab2} are only available in electronic form at the CDS via anonymous ftp to cdsarc.u-strasbg.fr (130.79.128.5) or via \url{http://cdsweb.u-strasbg.fr/cgi-bin/qcat?J/A+A/}}}

\author{V. Fallah Ramazani \inst{1} \and E. Lindfors \inst{1} \and K. Nilsson \inst{2}}

\institute{Tuorla Observatory, University of Turku, Väisäläntie 20, FI-21500 Piikkiö, Finland \and Finnish Centre for Astronomy with ESO (FINCA), University of Turku, Väisäläntie 20, FI-21500 Piikkiö, Finland}

\date{Received 26 January 2017 / Accepted 7 August 2017}

\abstract{
\textit{Aim:} We have collected the most complete multi-wavelength ($6.0 - 6.0 \times 10^{-18} cm$) dataset of very high energy (VHE) $\gamma$-ray emitting (TeV) BL Lacs, which are the most numerous extragalactic VHE sources. Using significant correlations between different bands, we aim to identify the best TeV BL Lac candidates that can be discovered by the current and next generation of imaging air Cherenkov telescopes.

\textit{Methods:} We formed five datasets from lower energy data, i.e. radio, mid-infrared, optical, X-rays, and GeV $\gamma$-ray, and five VHE $\gamma$-ray datasets to perform a correlation study between different bands and to construct the prediction method. The low energy datasets were averaged for individual sources, while the VHE $\gamma$-ray data were divided into subsets according to the flux state of the source.  We then looked for significant correlations and determined their best-fit parameters. Using the best-fit parameters we predicted the level of VHE $\gamma$-ray flux for a sample of 182 BL Lacs, which have not been detected at TeV energies. We identified the most promising TeV BL Lac candidates based on the predicted VHE $\gamma$-ray flux for each source.

\textit{Results:} We found 14 significant correlations between radio, mid-infrared, optical, $\gamma$-ray, and VHE $\gamma$-ray bands. The correlation between optical and VHE $\gamma$-ray luminosity is established for the first time. We attribute this to the more complete sample and more accurate handling of host galaxy flux in our work. We found nine BL Lac candidates whose predicted VHE $\gamma$-ray flux is high enough for detection in less than 25 hours with current imaging air Cherenkov telescopes.}

\maketitle
\section{Introduction}
\label{sec1}
The majority of extragalactic objects detected in the very high energy (VHE) $\gamma$-ray band ($>100$ GeV) are a rare type of radio-loud active galactic nuclei (AGN) called BL Lacs \citep{2011ApJ...739...73M}; \citep[][and references therein]{2009NIMPA.602...28D}. BL Lacs are characterized by large amplitude flux variability, highly polarized radiation in the radio and optical band, and apparent superluminal motion \citep{1978PhyS...17..265B}. The observed nuclear phenomenology of BL Lacs is interpreted as due to the presence of a relativistic jet. As a subclass of blazars, BL Lac jets are nearly aligned to the line of sight of the observer \citep{1979ApJ...232...34B}. Furthermore, BL Lac spectra in the optical band are rather featureless, and therefore the determination of the redshift is very challenging.

Very high energy $\gamma$-ray emitting (TeV) BL Lacs have an important role in physics and astronomy. As bright and distant objects, their VHE $\gamma$-ray spectra can be used to study the universe between us and BL Lacs (e.g. extragalactic background light (EBL) and intergalactic magnetic field (IGMF) \citep{2012Sci...338.1190A}). Moreover, their jets can be considered as natural particle acceleration laboratories.

BL Lacs have continuous two-humped shape spectral energy distribution (SED). The SED covers the electromagnetic spectrum from radio to VHE $\gamma$-ray and is dominated by non-thermal radiation. The first peak is located between the infrared and hard X-rays, while the second peak lies in the MeV-GeV part of the spectrum  \citep{1998MNRAS.301..451G}. There is also a claim that BL Lacs are counterparts of ultra high energy (PeV) neutrinos \citep[e.g.][]{2014MNRAS.443..474P}.

The first hump in the SEDs of BL Lacs is synchrotron emission from relativistic electrons spiralling in the magnetic field of the jet. There are competing explanations for the second hump. In the leptonic case, an inverse Compton (IC) mechanism describes the second hump of SED. The seed photons for the Compton scattering can be either external to the jet \citep{1989ApJ...340..162M, 1994ApJS...90..945D,1994ApJ...421..153S} or provided by the synchrotron emission of the jet itself (i.e. synchrotron self Compton; SSC \citep{1992MNRAS.258..657C,1992ApJ...397L...5M}). The most widely used source of external seed photons, the broad line region (BLR), which comprises line emitting clouds in rapid motion around the central black hole, is weak or absent in BL Lacs. Therefore, a single zone SSC model is typically used to model the second hump of the BL Lac SEDs, even though there is growing evidence that single zone models are not adequate \citep[e.g.][]{2014A&A...567A.135A}. In the hadronic scenario, the second hump of SED can be modelled using proton synchrotron emission \citep{2000NewA....5..377A, 2001APh....15..121M} or photopion production \citep{2014ApJ...797...89A}.
 
BL Lacs are categorized according to the frequency of the first peak of their SED ($\nu_{syn}$) \citep{1994MNRAS.268L..51G}. There are three types of BL Lacs: low peaked (LBL; $\nu_{syn}<10^{14}$ [Hz]), intermediate peaked (IBL; $10^{14}\leq\nu_{syn}<10^{15}$ [Hz]), and high peaked (HBL; $\nu_{syn}\geq10^{15}$ [Hz]) BL Lacs \citep{0004-637X-716-1-30}. The majority of TeV BL Lacs are HBLs.

Among the 1425 BL Lacs and BL Lac candidates in the fifth edition of the Roma-BZCAT catalogue {\citep{2015Ap&SS.357...75M}}, only 55 had been detected in the VHE $\gamma$-ray band by the end of December 2015. The TeV detected sample is observationally biased as no all-sky surveys from imaging air Cherenkov telescopes (IACTs) exists. This is due to the relatively long exposure time (typically 10 hours) needed to detect VHE $\gamma$-ray emission from typical BL Lacs combined with the size of the field of view of IACTs (three-five degrees). Because of these observational limitations, a prediction method to estimate VHE $\gamma$-ray fluxes of yet unobserved BL Lacs would be highly desirable.

In this paper, we have collected a complete multi-wavelength dataset of known TeV emitting BL Lacs and search for correlations between VHE $\gamma$-ray luminosity and the luminosity in lower energy bands. Previous works in this regard have been based on a small (<17) number of TeV BL Lacs. \citet{2002A&A...384...56C} selected 246 BL Lacs from five different samples and compared their X-rays (1 keV), radio (5 GHz), and optical (5500 \AA) properties with the known TeV BL Lacs. They found that all five TeV BL Lacs flux energy densities (hereafter flux) lie in two rectangles in the radio and X-ray plane as well as the optical and X-ray plane. In total, 33 TeV BL Lac candidates were introduced by these authors, of which $>65\%$  are now TeV BL Lac. \citet{2008MNRAS.385..119W} conducted a search to find correlations between lower frequency bands and the VHE $\gamma$-ray. Using a sample of 17 TeV BL Lacs, he investigated a correlation between X-ray and VHE $\gamma$-ray luminosities and found a coefficient of 0.76. No significant correlation was found between optical/radio and VHE $\gamma$-ray. \citet{2012RAA....12.1475F} investigated the correlation between radio and $\gamma$-ray photon flux, which was confirmed by a Spearman test on quasi-simultaneous data for 39 BL Lacs.

This paper aims at introducing a new TeV BL Lac candidate list based on correlations found in TeV detected BL Lacs. Two samples of BL Lacs are collected here: one TeV detected BL Lac sample and another non-TeV detected BL Lac sample which are presented in Section \ref{sec2}. Section \ref{sec3} contains our cross band correlation study. Using the significant correlations we found, Section \ref{sec4} presents a prediction of the TeV flux for non-TeV BL Lacs.  Finally, Section \ref{sec5} contains the discussion and conclusions of this paper.

\section{Samples and data}
\label{sec2}
Two samples of BL Lac objects are used in this study, a TeV BL Lac sample, which is used to build the prediction method, and a non-TeV BL Lac sample to which the prediction method is applied in order to find the best TeV BL Lac candidates.

The TeV BL Lac sample is based on the list in TeVCAT\footnote{\url{http://tevcat.uchicago.edu}}. There were 55 BL Lacs listed in the TeVCAT online catalogue at the end of 2015, but more TeV BL Lacs have been discovered since then (see discussion in Section \ref{sec4}). Among the BL Lacs listed there, IC310 and HESS~J1943+213 have uncertain classification \citep{2015arXiv150203559S,2015arXiv150203016K}. Additionally, the VHE $\gamma$-ray spectral properties of six TeV BL Lacs have not been published yet\footnote{The omitted sources from TeV sample are S2~0109+22, RX~J1136.8+6737, RGB~J0136+391,   MS~1221.8+2452, S3~1227+25, and RBS~0723.}. Therefore, our TeV BL Lac sample contains 47 BL Lacs. The sample is presented in Table \ref{tab1}, which is an example of the complete online version.

Six bands in the electromagnetic spectrum were selected for the correlation study based on the availability of the data. In the radio band, we selected 4.85 GHz. The mid-infrared band ($ 6.44 \times 10^{4}$ GHz) was selected based on the availability data from Wide-field Infrared Survey Explorer (\textit{WISE}). In the optical, the R band (Cousins; $4.85 \times 10^{5}$ GHz) was chosen based on the availability of long-term monitoring data. For the X-rays, a $2-10$ KeV ($4.83 \times 10^{8}- 2.41 \times 10^{9}$ GHz) range was selected and the range of $1-100$ GeV ($2.41 \times 10^{14}- 2.41 \times 10^{16}$ GHz) was selected in the $\gamma$-ray because the best sensitivity of the LAT instrument on board the \textit{FERMI} satellite lies in this range. Finally, a $>200$ GeV ($>4.83 \times 10^{16}$ GHz) range was selected for VHE $\gamma$-ray band. Subscripts R, I, O, X, $\gamma$ , and VHE were used as the indicators for radio, mid-infrared, optical, X-ray, $\gamma$-ray, and VHE $\gamma$-ray bands, respectively.

The fifth edition of the Roma-BZ catalogue \citep{2015Ap&SS.357...75M} contains 1425 BL Lacs and BL Lac candidates. Excluding TeV BL Lacs (55), BL Lacs with unknown (666) and uncertain (105) redshift leads to a sample of 599 objects. If we require flux information in at least three out of five lower energy bands, the number of non-TeV BL Lacs reduces to 182 objects (Table~\ref{tab2}).

In order to compute intrinsic properties, at least the redshift and multi-wavelength unabsorbed fluxes ($S^{obs.}=\nu f_{\nu}$ [$\rm   erg/cm^{2}/s$]) are needed. Additionally, the BL Lac type and photon index of local spectrum in the X-ray, $\gamma$-ray, and VHE $\gamma$-ray bands are needed for the K~correction of fluxes. However, owing to restricted data availability in the VHE $\gamma$-ray band, the unabsorbed energy flux density is substituted by measured energy flux density.

\subsection{Multi-wavelength flux data}
\label{mwl-f}
The radio fluxes at 4.85 GHz for both samples are collected from \citet{1999ApJ...525..127L,1998MNRAS.299..433F}; PMN, \citet{1995ApJS...97..347G}; 87GB, \citet{1991ApJS...75.1011G}; GB6, \citet{1994ApJS...90..173G,1996ApJS..103..427G}; and 1RXS, \citet{1999A&A...349..389V}. For the TeV sample, we were able to collect fluxes for all sources, whereas for 19 objects of the non-TeV sample we did not find the radio flux in the literature.  In the TeV sample the fluxes are between $3.88\times10^{-16}$ and ${2.93} \times10^{-13}$ [$\rm erg/cm^{2}/s$], and in non-TeV sample between $1.94\times10^{-16}$ and $1.81 \times10^{-13}$ [$\rm erg/cm^{2}/s$].

The mid-infrared fluxes for both samples are reproduced from \citet{2010AJ....140.1868W} based on the method described in \citet{2013ApJS..207...16M}. To estimate the mid-infrared average flux of the objects, we selected all the \textit{WISE} observations with S/N>2 in three bands (i.e. {(2.49, 6.51, and 8.82)} $\times 10^4$ GHz). The mid-infrared fluxes are between $8.50 \times 10^{-13} <S_{I}< 8.93 \times 10^{-11}$ and $2.65\times10^{-13} <S_{I}< 3.91\times10^{-10}$ [$\rm erg/cm^{2}/s$] for TeV and non-TeV samples, respectively.

The optical R-band fluxes for both samples are collected from the Tuorla blazar monitoring\footnote{\url{http://users.utu.fi/kani/1m/}} (Nilsson et al. 2017, submitted), \citep{2004A&A...427..387F} and \citep[USNO B1;][]{2003AJ....125..984M}.  We used average fluxes covering at least two years of observations. The optical fluxes are between $3.70 \times 10^{-13} < S_{O} < 1.08 \times 10^{-10}$ and $2.55 \times 10^{-14} < S_{O} < 5.65 \times 10^{-11}$ [$\rm erg/cm^{2}/s$] for the TeV and non-TeV samples, respectively. The flux values were corrected for the galactic extinction \citep{2011ApJ...737..103S} and the host galaxy contribution whenever the host galaxy parameters were available from good quality optical imaging \citep[e.g.][]{1999PASP..111.1223N,2003A&A...400...95N,2007A&A...475..199N}. The host galaxy contribution was integrated within an aperture corresponding to the aperture in the literature and subtracted from the R-band flux. Altogether 44 out of 47 objects in the TeV sample were treated in this way. For PKS~0447-439, BZB~J1010-3119, and PKS~1440-389, the host galaxy properties could not be found in the literature or determined from existing public data.  None of these three sources are categorized as host-galaxy-dominated BL Lacs (BZG) in \citet{2015Ap&SS.357...75M}. We also tested the host contribution using the result by \citet{2005ApJ...635..173S} that the BL Lac host galaxy luminosities are confined to a relatively narrow luminosity interval $M_R$ = -22.8 $\pm$ 0.5 [mag]. Assuming a further effective radius, $r_e = 10$ [kpc],  we computed the expected host galaxy flux within the aperture. For all three sources the host fraction was $<$25\%. Given that the host galaxy flux can be estimated with $\sim$50\% accuracy with this method, we did not subtract the host flux for these three sources.

We collected the X-ray fluxes of the TeV sample from the references listed in Table \ref{tab1}. For three sources (KUV~00311-1938, 1ES~1312-423, and MAGIC~J2001+435), we analysed the data (See \ref{2-1}). The X-ray fluxes of {91} targets in the non-TeV sample were collected from \citet{2013A&A...551A.142D}; we analysed the data for 33 of these sources (see \ref{2-1}) and reported these results in the literature for the first time. The X-ray fluxes are between $1.30 \times 10^{-12} <S_{X}< 5.31 \times 10^{-10}$ and $5.73 \times 10^{-14}<S_{X}<4.73 \times 10^{-11}$ [$\rm erg/cm^{2}/s$] for the TeV and non-TeV samples, respectively. Observations by \textit{SWIFT} are typically biased to flaring states as this instrument operates on a target of opportunity mode, but in our collection we also have data from pre-planned multi-wavelength observations. Additionally we averaged 5 to 10 fluxes for each source so that each flux would represent an average state rather than a flaring state.

The $\gamma$-ray flux (1-100 GeV) for both of the samples are collected from  \citep[3FGL;][]{2015ApJS..218...23A}, which provides average fluxes of the sources in the first four years of the {\textit Fermi } mission. The $\gamma$-ray flux varies between $1.90\rm\times10^{-11} <S_{\gamma}< 2.42\rm\times10^{-9}$ and $8.04\times10^{-12}<S_{\gamma}<2.64\rm\times10^{-9}$ [$\rm erg/cm^{2}/s$] for TeV and non-TeV samples, respectively.

We collected VHE $\gamma$-ray properties for the TeV sample from the references in Table \ref{tab1}. The pivot energies $E_t$, above which the integral flux is calculated, are highly inhomogeneous in the literature. Therefore, pivot energy ($E_t$), integral flux over the pivot energy ($f_{VHE(>E_t)}$), and best-fit spectral index ($\Gamma_{VHE}$) are collected to make the VHE $\gamma$-ray fluxes synchronized and comparable. We chose $E_t=200$ [GeV] as the homogenized pivot energy since it is the most frequent pivot energy of the collected data. We did not use EBL-corrected fluxes as these were not available for all sources. Since the sources are mostly located in relatively low redshift and detected in relatively low energies ($<500$ [GeV]), this should bias our study less than excluding sources for which there are no EBL corrected fluxes.

Some TeV BL Lacs have at least two VHE $\gamma$-ray integral fluxes over 200 GeV available in the literature. These BL Lacs are labelled as group~A while the rest are categorized as group~B. For group~A, we collected two flux values that represent the highest flux and lowest flux reported in the literature. For group~B, there either exists only one published measurement or there is no clear claim about integral flux variability.

\subsection{Redshifts and classification}

The redshifts for the non-TeV sample were collected from the Roma-BZCAT catalogue \citep{2015Ap&SS.357...75M}. For the TeV sample, we checked all redshifts one by one. In case of a discrepancy between reported redshifts in the literature, we used the most reliable value and noted the reason for using this value in Table \ref{tab1}.

The BL Lac type (HBL, IBL, and LBL) of the TeV sample and 67\% of the non-TeV sample were collected from \citet{2015ApJS..218...23A}. For the rest of the non-TeV sample, we used reported radio and X-ray fluxes from \citet{2015Ap&SS.357...75M} to compute the spectral slopes, $\alpha_{rx}$. We used $\alpha_{rx} = 0.75$ as the dividing value between HBLs and IBLs/LBLs \citep{2011A&A...529A.162H}.

\subsection{K~correction}

The fluxes were K~corrected with
\begin{equation}
S^{res} = S^{obs} \times (1-z)^{\alpha -1},
\end{equation}
where $\alpha$ is the energy spectral index in $S \propto \nu^{(-\alpha)}$ \citep{2013AJ....145...31K}. For the radio fluxes, we adopted $\alpha_R = 0.0$ for both samples \citep{2016RAA....16..103L}. The method described in \citet{2012ApJ...748...68D} was implemented to calculate the mid-infrared spectral index of all the objects in both samples. In the optical, we assumed $\alpha_O = 1.1$ for HBLs and $\alpha_O = 1.5$ for non-HBLs based on the average value given in \citet{2004A&A...419...25F} for both samples. For K~correction of the higher energy bands, we adopted
$\alpha = \alpha^{Ph} - 1,$ where $\alpha$ is the energy spectral index and $\alpha^{Ph}$ is the photon index \citep{2016RAA....16..103L}. The X-ray, $\gamma$-ray, and VHE $\gamma$-ray photon indices were collected from the same source of information described in Section \ref{mwl-f}. In the cases where the X-ray spectrum of the objects were described with log parabola
\begin{equation}
N(E) = {\frac{E}{E_{\rm Pivot}}}^{(-\alpha_{\rm Pivot} -\beta \log(\frac{E}{E_{\rm Pivot}}))},
\end{equation}
we used the $\alpha^{Ph}_{X} = \alpha_{Pivot} + (1.135 \times \beta)$ to calculate the equivalent power-law index. Here, $\beta$ is the curvature parameter of spectrum as described in \citet{2004A&A...413..489M}. Furthermore, we assumed $\alpha^{Ph}_{VHE}= 3.27$, which is the average of all the data points in our TeV sample, to calculate the predicted flux of non-TeV BL Lacs.
\subsection{SWIFT-XRT data}
\label{2-1} 
The multi-epoch event list obtained by the X-ray Telescope (XRT) \citep{2004SPIE.5165..201B} on board the \textit{SWIFT} satellite are processed using the HEASOFT package version 6.18. All the observations were performed in photon counting (PC) mode. We defined the source region as a circle of 20 pixels ($\sim$47") at the centre of the source, while we defined the background region by a ring centring at source with inner and outer radii of 40 ($\sim$94") and 80 pixels ($\sim$188"). We extracted the source and background spectra using \verb|XSELECT| task (v2.4c). If the source spectrum count rate exceeded 0.5 counts/s, we used a pile-up thread. To implement the pile-up thread, one pixel ($\sim  2.36''$) was excluded interactively from the source region, and the source and background regions were expanded by one pixel until the source spectrum count rate dropped below 0.5 counts/s. We used the \verb|xrtexpomap| task (v0.2.7) to correct the flux loss. This flux loss is due to not using some of the CCD pixels in data collection.  We used the \verb|xrtmkarf| task (v0.6.3) to take into account vignetting and bad pixels. Also the count loss caused by the pile-up was corrected by setting the PSF correction flag to "yes" and using the exposure map created by the \verb|xrtexpomap| task. The output of \verb|xrtmkarf| was saved in the ancillary response file for the next steps. We used the \verb|grppha| task to group source spectra in such a way that each bin contained 20 counts, thereby attaching the background, ancillary response file, and response matrix file to the source spectrum file. The flux and spectral parameters calculated employing the \verb|XSPEC| task (v12.9.0i). The photoelectric absorption coupled with the power-law model was fitted to the spectrum with fixed equivalent Galactic hydrogen column density ($n_H$). We obtained the $n_H$ values are from The Leiden/Argentine/Bonn (LAB) Survey of Galactic HI \citep{2005A&A...440..775K}. Then we averaged the model fluxes in the energy range of (2-10 keV) to make a single data point of X-ray flux for each source.

\section{Correlation study}
\label{sec3}

\begin{figure}
\begin{tikzpicture}
\begin{loglogaxis}[name =radio, ylabel={Log $L_{R}$ (erg/s)},
                   width=9cm, height=5.8cm, grid=major,
                   xticklabels={,,}, yticklabels={,40,41,42,43},
                   xmin=1e42, ymin=6e39, xmax=1e47, ymax=6e43]
\addplot[mark=o, blue, only marks] 
               table [x=VHE_L, y=RADIO,, col sep=comma] {Group_A.csv}; 
\addplot[mark=triangle, red, only marks] 
               table [x=VHE_H, y=RADIO,, col sep=comma] {Group_A.csv}; 
\addplot[mark=x, green, only marks] 
               table [x=VHE_SD, y=RADIO,, col sep=comma] {Group_B.csv};
\addplot[blue] [domain=1e42:1e47, samples=101,unbounded coords=jump]              
               {((1/3.2e17)*x)^(1/0.64)};
\end{loglogaxis}

\begin{loglogaxis}[name=IR, at=(radio.below south west),
                   anchor=north west, yshift = 0.24cm,
                   ylabel={Log $L_{I}$ (erg/s)},
                   width=9cm, height=5.8cm, grid=major,
                   xticklabels={,,}, yticklabels={42,43,44,45,46},
                   xmin=1e42, ymin=3e42, xmax=1e47, ymax=9e46]
\addplot[mark=o, blue, only marks] 
               table [x=VHE_L, y=IR,, col sep=comma] {Group_A.csv}; 
\addplot[mark=triangle, red, only marks] 
               table [x=VHE_H, y=IR,, col sep=comma] {Group_A.csv}; 
\addplot[mark=x, green, only marks] 
               table [x=VHE_SD, y=IR, col sep=comma] {Group_B.csv};
\addplot[blue] [domain=1e42:1e47, samples=101,unbounded coords=jump]
               {((1/6.3e11)*x)^(1/0.73)};
\end{loglogaxis}

\begin{loglogaxis}[name=optical, at=(IR.below south west),
                   anchor=north west, yshift = 0.24cm,
                   ylabel={Log $L_{O}$ (erg/s)},
                   width=9cm, height=5.8cm, grid=major,
                   xticklabels={,,}, yticklabels={42,43,44,45,46,47},
                   xmin=1e42, ymin=3e42, xmax=1e47, ymax=3e47]
\addplot[mark=o, blue, only marks] 
               table [x=VHE_L, y=OPTICAL,, col sep=comma] {Group_A.csv}; 
\addplot[mark=triangle, red, only marks] 
               table [x=VHE_H, y=OPTICAL,, col sep=comma] {Group_A.csv}; 
\addplot[mark=x, green, only marks] 
               table [x=VHE_SD, y=OPTICAL, col sep=comma] {Group_B.csv};
\addplot[blue] [domain=1e42:1e47, samples=101,unbounded coords=jump]
               {((1/8.71e6)*x)^(1/0.83)};
\addplot[blue, dashed] [domain=1e42:1e47, samples=101,
                        unbounded coords=jump]
               {((1/8.13e3)*x)^(1/0.90)};
\addplot[red] [domain=1e42:1e47, samples=101,unbounded coords=jump]
               {((1/3.2e12)*x)^(1/0.72)};
\end{loglogaxis}

\begin{loglogaxis}[name=xray, at=(optical.below south west),
                   anchor=north west, yshift = 0.24cm,
                   ylabel={Log $L_{X}$ (erg/s)},
                   width=9cm, height=5.31cm, grid=major,
                   xticklabels={,,}, yticklabels={,43,44,45,46},
                   xmin=1e42, ymin=5e42, xmax=1e47, ymax=5e46]
\addplot[mark=o, blue, only marks] 
               table [x=VHE_L, y=X_RAY,, col sep=comma] {Group_A.csv}; 
\addplot[mark=triangle, red, only marks] 
               table [x=VHE_H, y=X_RAY,, col sep=comma] {Group_A.csv}; 
\addplot[mark=x, green, only marks] 
               table [x=VHE_SD, y=X_RAY,, col sep=comma] {Group_B.csv};
\end{loglogaxis}

\begin{loglogaxis}[name=gamma, at=(xray.below south west),
                   anchor=north west, yshift = 0.24cm,
                   ylabel={Log $L_{\gamma}$ (erg/s)},
                   xlabel={Log $L_{VHE}$ (erg/s)},
                   width=9cm, height=5.8cm, grid=major,
                   xticklabels={,43,44,45,46,47},
                   yticklabels={,45,46,47,48},
                   xmin=1e42, ymin=1.1e44, xmax=1e47, ymax=1e49]
\addplot[mark=o, blue, only marks] 
               table [x=VHE_L, y=GAMMA,, col sep=comma] {Group_A.csv}; 
\addplot[mark=triangle, red, only marks] 
               table [x=VHE_H, y=GAMMA,, col sep=comma] {Group_A.csv}; 
\addplot[mark=x, green, only marks] 
               table [x=VHE_SD, y=GAMMA,, col sep=comma] {Group_B.csv};
\addplot[red] [domain=1e42:1e47, samples=101,unbounded coords=jump]
               {((1/1.5e9)*x)^(1/0.77)};
\addplot[blue] [domain=1e42:1e47, samples=101,unbounded coords=jump] 
               {((1/3.55e4)*x)^(1/0.86)};
\addplot[blue, dashed] [domain=1e42:1e47, samples=101,
                        unbounded coords=jump]
               {((1/4.27e1)*x)^(1/0.92)};
\end{loglogaxis}

\node at (0.5cm,3.6cm) {(a)};
\node at (0.5cm,-0.7cm) {(b)};
\node at (0.5cm,-5.2cm) {(c)};
\node at (0.5cm,-9.5cm) {(d)};
\node at (0.5cm,-12.8cm) {(e)};
\end{tikzpicture}
\caption[TeV Sample distribution in L-L planes]
{\label{Fig_corrs} Luminosity in the VHE $\gamma$-ray band vs. luminosity in radio (a), mid-infrared (b), optical (c), X-ray   (d), and $\gamma$-ray (e) bands in logarithmic scale. The different symbols represent the data of different groups (see text): low state group A (\textcolor{blue}{circle}), high state group A (\textcolor{red}{triangle}) and group B (\textcolor{green}{cross}). The correlation functions of the various groups are shown with low state group A (\textcolor{blue}{blue dashed line}), the combined dataset of high state group A and group B (\textcolor{red}{red line}), and the combined dataset of low state group A and group B (\textcolor{blue}{blue line}).}
\end{figure}
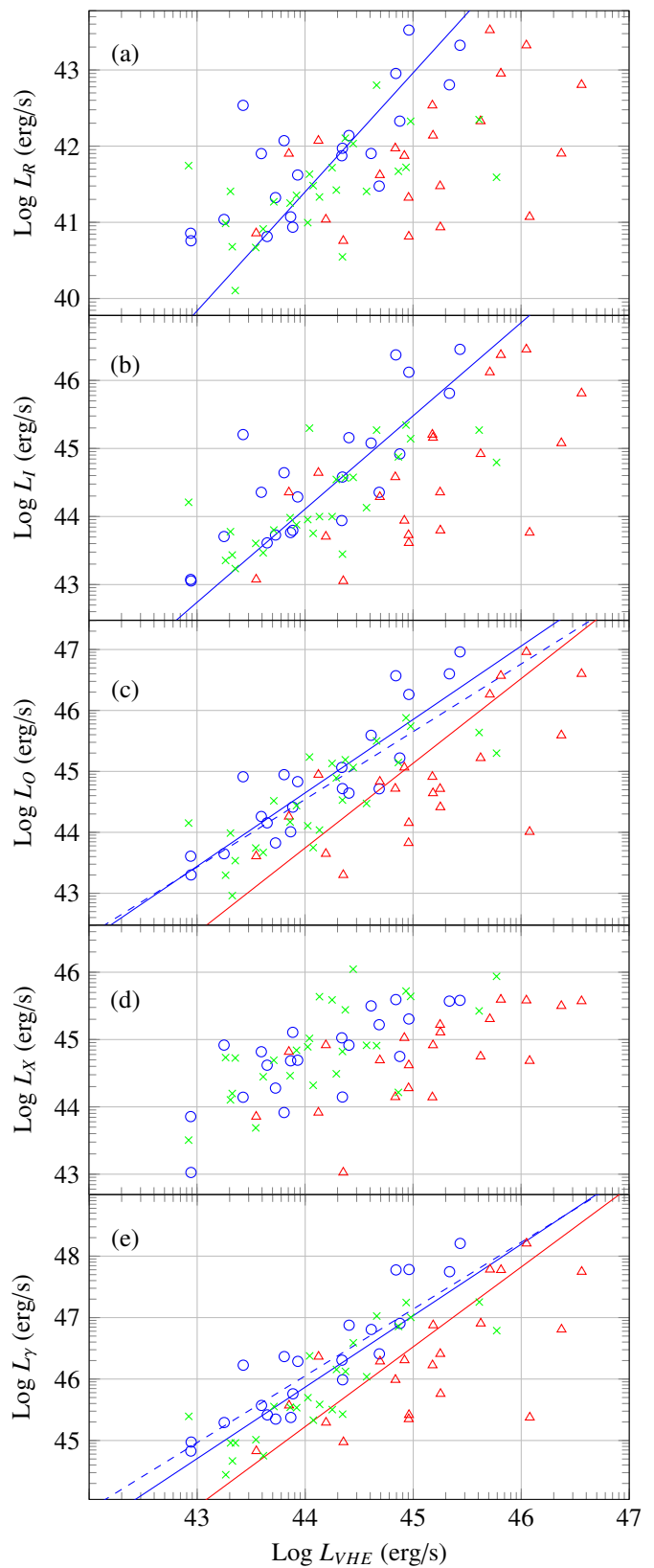

We calculated the luminosities in lower energy bands (i.e. radio, mid-infrared, optical, X-ray, and $\gamma$-ray) with the unabsorbed K~corrected flux in each band assuming the $\Lambda$CDM model for a flat universe, $H_{0}=67.3$ [$\rm km/s/Mpc$] and $\Omega_{M}=0.315$ \citep{2014A&A...571A..16P}, using the NED Cosmology Calculator-I\footnote{\url{http://www.astro.ucla.edu/~wright/CosmoCalc.html}} \citep{2006PASP..118.1711W}.
For the TeV sample, VHE $\gamma$-ray luminosities were calculated from the K~corrected flux and categorized into single TeV-detected (group~B, 26 objects) and multiple TeV-detected (group~A, 21 objects) groups. Figure \ref{Fig_corrs} illustrates the distribution of TeV sample in different luminosity-luminosity planes. Group~A is further divided into two subclasses that represent the high and low states of the sources in VHE $\gamma$-ray band. One VHE $\gamma$-ray dataset is formed based on group~B. Two VHE $\gamma$-ray datasets are constructed based on the two subclasses of group~A. Furthermore, two additional datasets are formed by combining single-detected TeV BL Lacs and each subclasses of multiple detected TeV BL Lacs. In total, there are 10 datasets, including 5 in lower energy bands and 5 in the VHE $\gamma$-ray band. These 10 datasets form 35 pairs.

We used the non-parametric partial Kendall~$\tau$ rank correlation test \citep{1996MNRAS.278..919A} to test for a possible correlation, and the significance of this correlation, between each pair
of datasets, while taking the effect of redshift on each luminosity-luminosity correlation into account. The null hypothesis probability was set at $p_{value}\le  1.5 \times 10^{-3}$, which gives the false alarm probability of 5\% with the number of tested correlations in our sample. Table \ref{tab3} shows the results of our correlation study. The first two columns show which datasets are considered as correlation pairs. Columns 3 to 7 show the number of data points, partial Pearson coefficient, Kendall $\tau_b$ coefficient, correlation significance, and the null hypothesis probability, respectively. Fourteen pairs of datasets have significant correlations, among these pairs 8 pairs have VHE $\gamma$-ray band as the second parameter. The number of significant correlations (14) clearly exceeds the expected number from chance only ($\sim$ 1) and from this point on we treat all of these as genuine correlations. Moreover, to measure the dispersion of the data with respect to the best-fit line, we calculated the partial Pearson coefficient (PPC) in logarithmic space for each luminosity-luminosity correlation (table \ref{tab3}, column 3). The PPC gives an approximation of the data dispersion with respect to the best-fit line in logarithmic space. Owing to restrictions, which are implied from the number of data points together with the PCC assumptions, we adopted the partial Kendall coefficient to test the existence of correlation between each pair of datasets.

For the 14 pairs of datasets that show significant correlation, we used the bi-sectional ordinary least-squares (OLS) regression method \citep{1990ApJ...364..104I} to fit two models: (1) a linear model ($Y = aX + b$) and (2) a power-law model ($Y = 10^b \times X^a$). The latter was made by fitting a linear model in log-log space. The sum of squared residuals (SSD) was used to judge which of these two models can describe the correlation better. The error bars of the model parameters were calculated applying a standard bootstrapping method \citep{tEFR93a}. We checked the model dependency on outliers by comparing the calculated error bars from the bootstrapping method with those from the Jackknife method \citep{tEFR81a}. All the significant correlations can be described better with the power-law model and are independent of outlier data points. Table \ref{tab4} shows the best-fit model parameters for all the significant correlations together with the number of times ($N_p$) that the correlations produced luminosity values closest to the median of the VHE $\gamma$-ray predicted luminosity level (see section \ref{sec4}). The significant correlations are also illustrated in figure~\ref{Fig_corrs}.

\subsection{Significant correlations}
\label{sigcor}
We find 14 significant correlations ($p_{value}\le  1.5 \times 10^{-3}$; table \ref{tab4}), and within these correlations six do not include the VHE $\gamma$-ray band\footnote{Lower energy bands correlations are listed in the lower part of table \ref{tab4}.}.

\textit{Radio versus mid-infrared:} This correlation is expected as the radio and mid-infrared emission should originate by synchrotron radiation from the same population of relativistic electrons. However, to our knowledge this is the first confirmation of this correlation in the literature.

\textit{Radio versus optical:} This correlation is another confirmation of the similarity of the emission regions at these two wavelengths. \citet{2016A&A...593A..98L} estimated that 10-50\% of the optical emission of TeV BL Lacs originates from the same region as the radio emission. However, their sample only contained northern sky TeV BL Lacs (32 objects), here we establish the same correlation for a larger sample.

\begin{table}[ht]
\begin{center}
\caption{Cross band correlations study results} 
\label{tab3}      
\addtolength{\tabcolsep}{-0.1pt}         
\begin{tabular}{ccccccc}
\hline
\hline
\multicolumn{2}{c}{Parameters}&\multirow{2}{*}{N}&\multirow{2}{*}{PPC\tablefootmark{$\dagger$}}&\multirow{2}{*}{$\tau_b$\tablefootmark{$\ddagger$}}&\multirow{2}{*}{$\sigma$\tablefootmark{$\star$}}&\multirow{2}{*}{\textit{$p_{value}$}\tablefootmark{$\ast$}}\\
X & Y& && & \\
\hline
$L_{R}$&	 $L_{VHE,(H+SD)}$&	 47&	 0.38&	 0.31&	 3.03&	 $2.4 \times 10^{-3}$\\
$L_{R}$&	 $L_{VHE,(L+SD)}$&	 47&	 0.24&	 0.32&	 3.18&	 $1.5 \times 10^{-3}$\\
$L_{R}$&	 $L_{VHE,(H)}$	&	 21&	 0.13&	 0.17&	 1.04&	 $3.0 \times 10^{-1}$\\
$L_{R}$&	 $L_{VHE,(L)}$	&	 21&	 0.37&	 0.36&	 2.23&	 $2.6 \times 10^{-2}$\\
$L_{R}$&	 $L_{VHE,(SD)}$	&	 26&	 0.12&	 0.20&	 1.41&	 $1.6 \times 10^{-1}$\\
$L_{I}$&	 $L_{VHE,(H+SD)}$&	 47&	 0.25&	 0.25&	 2.49&	 $1.3 \times 10^{-2}$\\
$L_{I}$&	 $L_{VHE,(L+SD)}$&	 47&	 0.32&	 0.34&	 3.31&	 $9.3 \times 10^{-4}$\\
$L_{I}$&	 $L_{VHE,(H)}$	&	 21&	 0.25&	 0.28&	 1.74&	 $8.3 \times 10^{-2}$\\
$L_{I}$&	 $L_{VHE,(L)}$	&	 21&	 0.44&	 0.43&	 2.62&	 $8.7 \times 10^{-3}$\\
$L_{I}$&	 $L_{VHE,(SD)}$	&	 26&	 0.28&	 0.27&	 1.86&	 $6.3 \times 10^{-2}$\\
$L_{O}$&	 $L_{VHE,(H+SD)}$&	 47&	 0.56&	 0.39&	 3.86&	 $1.2 \times 10^{-4}$\\
$L_{O}$&	 $L_{VHE,(L+SD)}$&	 47&	 0.48&	 0.46&	 4.51&	 $6.4 \times 10^{-6}$\\
$L_{O}$&	 $L_{VHE,(H)}$	&	 21&	 0.47&	 0.34&	 2.10&	 $3.6 \times 10^{-2}$\\
$L_{O}$&	 $L_{VHE,(L)}$	&	 21&	 0.66&	 0.53&	 3.29&	 $1.0 \times 10^{-3}$\\
$L_{O}$&	 $L_{VHE,(SD)}$	&	 26&	 0.30&	 0.24&	 1.71&	 $8.8 \times 10^{-2}$\\
$L_{X}$&	 $L_{VHE,(H+SD)}$&	 47&	 0.25&	 0.24&	 2.32&	 $2.0 \times 10^{-2}$\\
$L_{X}$&	 $L_{VHE,(L+SD)}$&	 47&	 0.39&	 0.28&	 2.72&	 $6.5 \times 10^{-3}$\\
$L_{X}$&	 $L_{VHE,(H)}$	&	 21&	 0.49&	 0.42&	 2.59&	 $9.7 \times 10^{-3}$\\
$L_{X}$&	 $L_{VHE,(L)}$	&	 21&	 0.59&	 0.41&	 2.55&	 $1.1 \times 10^{-2}$\\
$L_{X}$&	 $L_{VHE,(SD)}$	&	 26&	 0.12&	 0.14&	 0.97&	 $3.3 \times 10^{-1}$\\
$L_{\gamma}$&	 $L_{VHE,(H+SD)}$&	 47&	 0.67&	 0.46&	 4.54&	 $5.7 \times 10^{-6}$\\
$L_{\gamma}$&	 $L_{VHE,(L+SD)}$&	 47&	 0.57&	 0.50&	 4.90&	 $9.5 \times 10^{-7}$\\
$L_{\gamma}$&	 $L_{VHE,(H)}$	&	 21&	 0.47&	 0.36&	 2.22&	 $2.6 \times 10^{-2}$\\
$L_{\gamma}$&	 $L_{VHE,(L)}$	&	 21&	 0.73&	 0.62&	 3.84&	 $1.2 \times 10^{-4}$\\
$L_{\gamma}$&	 $L_{VHE,(SD)}$	&	 26&	 0.48&	 0.36&	 2.49&	 $1.3 \times 10^{-2}$\\
\hline

$L_{R}$&	 $L_{I}$	&	 47&	 0.63&	 0.48&	 4.73&	 $2.2 \times 10^{-6}$\\
$L_{R}$&	 $L_{O}$	&	 47&	 0.73&	 0.54&	 5.25&	 $1.5 \times 10^{-7}$\\
$L_{R}$&	 $L_{X}$	&	 47&	 -0.03&	 0.08&	 0.81&	 $4.2 \times 10^{-1}$\\
$L_{R}$&	 $L_{\gamma}$	&	 47&	 0.74&	 0.55&	 5.44&	 $5.4 \times 10^{-8}$\\
$L_{I}$&	 $L_{O}$	&	 47&	 0.74&	 0.57&	 5.59&	 $2.2 \times 10^{-8}$\\
$L_{I}$&	 $L_{X}$	&	 47&	 0.20&	 0.19&	 1.90&	 $5.8 \times 10^{-2}$\\
$L_{I}$&	 $L_{\gamma}$	&	 47&	 0.71&	 0.57&	 5.55&	 $2.9 \times 10^{-8}$\\
$L_{O}$&	 $L_{X}$	&	 47&	 0.19&	 0.20&	 1.98&	 $4.7 \times 10^{-2}$\\
$L_{O}$&	 $L_{\gamma}$	&	 47&	 0.86&	 0.72&	 7.01&	 $2.3 \times 10^{-12}$\\
$L_{X}$&	 $L_{\gamma}$	&	 47&	 0.10&	 0.20&	 1.91&	 $5.6 \times 10^{-2}$\\
\hline
\end{tabular}
\tablefoot{
\tablefoottext{$\dagger$}{Partial Pearson coefficient.}\tablefoottext{$\ddagger$}{Partial Kendall coefficient.}\tablefoottext{$\star$}{Correlation significant based on $\tau_b$.}\tablefoottext{$\ast$}{Null hypothesis probability.}}
\end{center}
\end{table}

\textit{Radio versus $\gamma$-ray:} As discussed, for example in \citet{2012Sci...338.1445N}, radio to $\gamma$-ray correlation simply reflects the connection between the kinetic power of the jet and the energy dissipation, and this correlation is therefore expected. A correlation between these two bands has also been found in \citep[e.g.][]{2009ApJ...696L..17K,2011A&A...535A..69N} for larger samples of blazars.

\textit{Mid-infrared versus optical:} As the two bands are rather close in energy, a correlation between these two bands is expected when the host galaxy flux is correctly subtracted from the optical data because the host galaxy dilutes the optical band much more; this is correlation is found this work.

{\textit{Mid-infrared versus $\gamma$-ray}: \citet{2016ApJ...827...67M} found a strong connection between the fluxes and spectral properties between these two bands considering a large sample of \textit{Fermi}--LAT blazars. Therefore, it is not surprising that we find this correlation in our study as well.}

\textit{Optical versus $\gamma$-ray:} This correlation was also found for a larger sample of blazars in \citep{2014MNRAS.439..690H}. It is expected as the optical photons serve as seed photons for IC emission in the $\gamma$-ray band, but also because the optical and gamma-ray emission are produced by the same electrons.

\textit{Radio versus $VHE_{L+SD}$:} This correlation was not seen in the earlier work \citep{2008MNRAS.385..119W}, probably because of a significantly smaller sample size. The correlation is rather natural extension to observed radio versus $\gamma$-ray correlation.

{\textit{Mid-infrared versus $VHE_{L+SD}$:} Similar to the radio versus $VHE_{L+SD}$ correlation, this correlation is also a rather natural extension to the observed mid-infrared versus $\gamma$-ray correlation. The found correlation is in agreement with \citet{2013ApJS..206...13M} using mid-infrared to select new TeV blazar candidates.}

\textit{Optical versus $VHE_{L/L+SD/H+SD}$:} The \textit{MAGIC} Collaboration has been successfully using optical monitoring data to trigger VHE $\gamma$-ray observations \citep[e.g.][and references therein]{2016MNRAS.459.3271A,2015MNRAS.451..739A,2012A&A...544A.142A} and therefore the connection between the two wavebands in not unexpected. However, in previous similar works \citep[e.g.][]{2002A&A...384...56C,2008MNRAS.385..119W} this correlation was not significant. Comparing their datasets to work presented here, we note that in previous works the host galaxy fluxes were not subtracted from optical fluxes and this possibly led to non-significant correlation in these studies.

\textit{$\gamma$-ray versus $VHE_{L+SD/H+SD}$:} It was expected that the number of significant correlations between $\gamma$-ray and VHE $\gamma$-ray would be higher than the other bands because the bands are close in energy and the spectra in the two bands typically connect smoothly. However, the VHE $\gamma$-ray single-detected dataset does not show any significant correlation to $\gamma$-ray. This can be due to the non-simultaneous data in the two bands because the single-detected VHE $\gamma$-ray data are mostly from the flaring state, while our $\gamma$-ray dataset is a long-term average, which does not reflect the flaring activity. Additionally, the number of data points could also be too small because the combined datasets show significant correlations.

\textit{$\gamma$-ray versus $VHE_{L}$:} The long-term average $\gamma$-ray dataset mostly represents the quiescent state of sources, which is also present in the $VHE_{L}$ datasets.

\begin{table}[ht]
\begin{center}
\caption{\label{tab4}Power law model parameters for significant correlations}                         
\addtolength{\tabcolsep}{-0.7pt}
\begin{tabular}{lccccc}
\hline
\hline
\multirow{2}{*}{No.}&\multicolumn{2}{c}{Parameters}&\multicolumn{2}{c}{Model parameters\tablefootmark{$\dagger$}}&\multirow{2}{*}{$N_P$\tablefootmark{$\ddagger$}}\\
&X & Y& a& b&\\
\hline
$c_1$&$L_{R}$&	 $L_{VHE,(L+SD)}$&	$ 0.64 \pm 0.09 $&	$ 17.50 \pm 3.77$&48\\
$c_2$&$L_{I}$&	 $L_{VHE,(L+SD)}$&	$ 0.73 \pm 0.06 $&	$ 11.80 \pm 2.65$&59\\
$c_3$&$L_{O}$&	 $L_{VHE,(L+SD)}$&	$ 0.83 \pm 0.04 $&	$ 6.94 \pm 1.82$&44\\
$c_4$&$L_{O}$&	 $L_{VHE,(L)}$	&	$ 0.90 \pm 0.05 $&	$ 3.91 \pm 2.32$&31\\
$c_5$&$L_{O}$&	 $L_{VHE,(H+SD)}$&	$ 0.72 \pm 0.08 $&	$ 12.50 \pm 3.65$&41\\
$c_6$&$L_{\gamma}$&	 $L_{VHE,(L+SD)}$&	$ 0.86 \pm 0.03 $&	$ 4.55 \pm 1.58$&52\\
$c_7$&$L_{\gamma}$&	 $L_{VHE,(L)}$	&	$ 0.92 \pm 0.04 $&	$ 1.63 \pm 1.69$&24\\
$c_8$&$L_{\gamma}$&	 $L_{VHE,(H+SD)}$&	$ 0.77 \pm 0.07 $&	$ 9.17 \pm 3.38$&34\\
\hline
$c_9$&$L_{R}$&	 $L_{I}$	&	$ 0.81 \pm 0.06 $&	$ 10.80 \pm 2.32$&\\
$c_{10}$&$L_{R}$&	 $L_{O}$	&	$ 0.85 \pm 0.04 $&	$ 9.09 \pm 1.56$&\\
$c_{11}$&$L_{R}$&	 $L_{\gamma}$	&	$ 0.86 \pm 0.04 $&	$ 10.20 \pm 1.45$&\\
$c_{12}$&$L_{I}$&	 $L_{O}$	&	$ 0.89 \pm 0.03 $&	$ 5.13 \pm 1.37$&\\
$c_{13}$&$L_{I}$&	 $L_{\gamma}$	&	$ 0.88 \pm 0.04 $&	$ 6.95 \pm 1.58$&\\
$c_{14}$&$L_{O}$&	 $L_{\gamma}$	&	$ 0.95 \pm 0.02 $&	$ 3.55 \pm 0.93$&\\
\hline
\end{tabular}
\tablefoot{
\tablefoottext{$\dagger$}{Power law model parameters, the function is in the form of $Y = 10^b \times X^a$.}\tablefoottext{$\ddagger$}{$N_P$: Number of times that the correlation produced luminosity value closest to the median of $L_{VHE,pred.}$ (see section \ref{sec4})}}
\end{center}
\end{table}

\subsection{Non-significant correlations}
\label{nonsig}
Out of 35 correlations tested, 21 were not significant. Two particular cases call for a closer look: 

\begin{itemize}
\item \textit{X-ray versus VHE $\gamma$-ray correlations:} We did not find any significant correlations between X-ray and VHE $\gamma$-ray bands, which is rather surprising in light of the previous studies. In previous works, a correlation between the two bands has been seen(\citet{2008MNRAS.385..119W}). There are well-known correlations between these two bands for single sources during their flaring activity (e.g. \citet{2011ApJ...738...25A}) and low states (e.g. \citet{2015A&A...576A.126A}). However, in our case the data in these two bands, where the BL Lac sources typically show the largest amplitude of variability, are non-simultaneous. The correlations containing the $VHE_{L}$ fluxes have p-values only marginally below the set p-value, which is probably due to X-ray sample representing the low state rather than the high state, but not consisting only of low state fluxes.

\item \textit{$\gamma$-ray versus $VHE_{H}$ correlation:} no significant correlation was detected between $\gamma$-ray versus $VHE_{H}$ datasets. In most of our sources the second peak of the SED is located in the 10-100\,[GeV] band, i.e. the $\gamma$-ray band is located below the $\gamma_{break}$ while VHE band $\gamma$-ray located above $\gamma_{break}$. Typically, the variability has larger amplitude above $\gamma_{break}$, so in our case at the VHE $\gamma$-ray band. Therefore the result is not totally surprising, in particular as our $VHE_{H}$ dataset contains the highest ever detected VHE $\gamma$-ray fluxes while the $\gamma$-ray dataset is averaged. This can also be distinguished visually from figure \ref{Fig_corrs}.
\end{itemize}  

\section{TeV BL Lac candidates}
\label{sec4}

Based on the significant correlations found in the TeV sample we formed {eight }prediction functions, $\log (L_{VHE,pred.}) = b + a\log(L_{R/ I/\rm O/X/\gamma})$. These functions were then applied to the lower energy band luminosities of non-TeV BL Lacs to predict the level of VHE $\gamma$-ray luminosity over 200 GeV including its $1\sigma$ error bar. In the optical, we subtracted the host galaxy contribution assuming again $M_R = -22.8$ and $r_e$ = 10 kpc prior to applying the optical-VHE correlation.
Depending on the availability of the low energy data, at least seven levels of VHE $\gamma$-ray luminosity are calculated for each non-TeV BL Lac. The median of $L_{VHE,pred.}$ for each source is considered as the predicted level of VHE $\gamma$-ray luminosity. The multi-wavelength fluxes together with minimum, median and maximum of $S_{VHE,pred.}$ are presented in table \ref{tab2} for non-TeV sample. Moreover, the last column indicates which correlations, listed in table \ref{tab4}, produced the closest value to the median of $S_{VHE,pred.}$. The correlation between $L_I$ and $VHE_{L+SD}$ ($c_2$) produced the closest value to the median of $S_{VHE,pred.}$ , which is more than the other correlations. However, the two correlations that most frequently produce the median flux $S_{VHE,pred.}$ are $c_5$ and $c_6$ ($L_O$ versus $L_{VHE,(H+SD)}$ and $L_{\gamma}$ versus $L_{VHE,(L+SD)}$). It is also clear that all the correlations are used to calculate the median flux $S_{VHE,pred.}$ multiple times and therefore none of the correlations can be excluded from the study.

We compared these predicted fluxes with the sensitivity of current generation of IACTs, using the low zenith distance integral sensitivity above 200 GeV of \textit{MAGIC} telescopes\footnote{{The \textit{MAGIC} sensitivity is obtained from 50 hours observation of Crab Nebula as VHE $\gamma$-ray standard candle. Therefore, it depends on the observed spectrum of Crab Nebula.}} \citep{2016APh....72...76A}. This shows that 53 non-TeV BL Lacs in our sample are expected to be detectable with the current generation of IACTs. Hereafter, we labelled these 53 sources TeV BL Lac  candidates. To give a ``short-list'' of the best candidates, we also  list in table \ref{tab2} the nine most promising TeV candidates (see below); a complete table of the non-TeV BL Lacs is available online.
 
Considering only the minimum predicted luminosity, six of our candidates should be detectable within 12 hours of observations and three should be detectable within 12-25 hours; we call these nine objects the most promising candidates (see above). A further four sources should be detectable within 25-50 hours. In total 27 candidates have their minimum predicted fluxes above the faintest VHE $\gamma$-ray flux detected by current IACTs \citep[the lowest state of 1ES~0229+200;][]{2015ICRC...34..762C}).

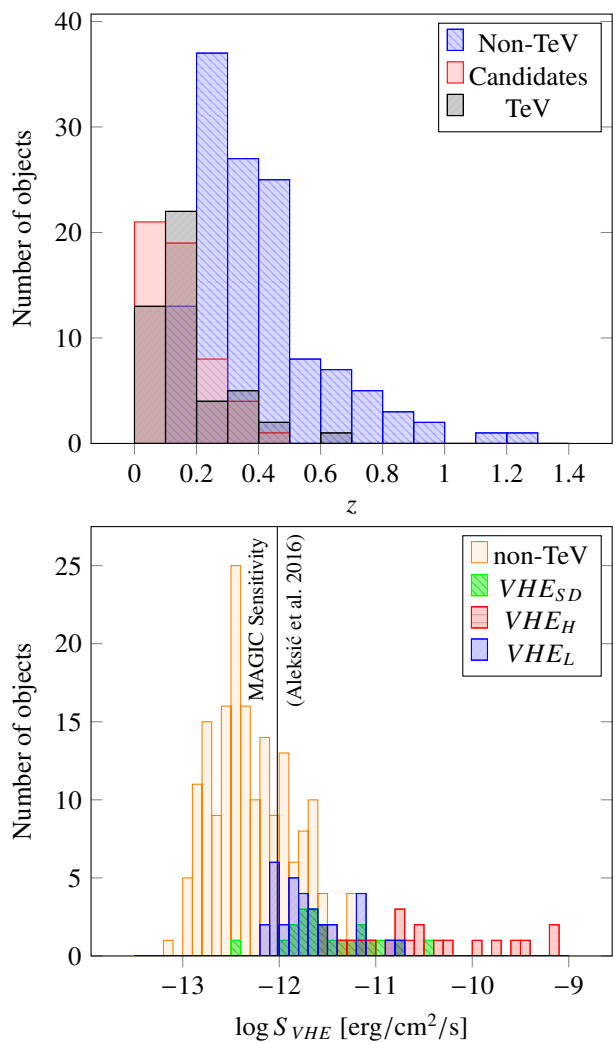
\begin{figure}
\begin{tikzpicture}
\begin{axis}[ybar, ymin=0, xlabel={\textit{z}},
             ylabel={Number of objects}, xtick pos=left]
\addplot +[fill opacity=.50,
           postaction={pattern color=blue,pattern=north west lines},
           hist={bins=14, data min=0.0, data max=1.4}]
             table [y index=5,col sep=comma]{redshifts.csv};

\addplot +[fill opacity=.50,
           hist={bins=14, data min=0.0, data max=1.4}] 
             table [y index=4,col sep=comma]{redshifts.csv};

\addplot +[fill opacity=.2,color=black,
           postaction={pattern=north east lines},
           hist={bins=14, data min=0.0, data max=1.4}]
              table [y index=0,col sep=comma]{redshifts.csv};
           
\legend{Non-TeV,Candidates, TeV}
\end{axis}
\end{tikzpicture}

\begin{tikzpicture}
\begin{axis}[ybar, ymin=0, xlabel={$\log S_{VHE}$ [$\rm erg/cm^2/s$]},
             ylabel={Number of objects},xtick pos=left]          
\addplot +[fill opacity=.1,color=orange,
           hist={bins=45, data min=-13.5, data max=-9}] 
           table [y index=7,col sep=comma]{fdist.csv};

\addplot +[fill opacity=.50,color=green,
          postaction={pattern=north west lines},
          hist={bins=45, data min=-13.5, data max=-9}] 
           table [y index=6,col sep=comma]{fdist.csv};

\addplot +[fill opacity=.20,color=red,
           postaction={pattern=horizontal lines},
           hist={bins=45, data min=-13.5, data max=-9}] 
           table [y index=5,col sep=comma]{fdist.csv};

\addplot +[fill opacity=.2,color=blue,
           postaction={pattern=north east lines},
           hist={bins=45, data min=-13.5, data max=-9}] 
           table [y index=4,col sep=comma]{fdist.csv};
           
\draw [black] (axis cs:-12.02,0) -- (axis cs:-12.02,30) node [rotate=90, pos=0.72, above] {{\tiny MAGIC Sensitivity}};

\draw [black] (axis cs:-9.,0) -- (axis cs:-13.5,0);

\node at (axis cs:-12.02,21.5) [rotate=90, below] {{\tiny (Aleksi{\'c} et al. 2016)}};

\legend{non-TeV, $VHE_{SD}$,$VHE_{H}$,$VHE_{L}$}
\end{axis}
\end{tikzpicture}
\caption{\label{histo} {\em Upper panel}: Redshift distribution of TeV BL Lacs, TeV Candidates, and non-TeV BL Lacs. {\em Lower panel}: Predicted VHE gamma-ray fluxes of non-TeV BL Lacs compared to three different sub-samples of known VHE emitting BL Lacs: single-detected sources (SD), high VHE flux state (H), and low VHE flux state (L).}
\end{figure}

In Fig. \ref{histo}, we compare our candidate sample to sub-samples of known VHE gamma-ray emitting BL Lacs:\ the redshift distributions of previously known TeV BL Lacs, the non-TeV BL Lac sample studied here, and our 53 candidates (figure \ref{histo}, upper panel). This comparison shows that redshift distribution of the candidates is similar to the distribution of the known BL Lacs. One should note that the EBL absorption was not taken into account for the predicted level of VHE $\gamma$-ray emission. However, the best candidates are at low redshift and therefore the affect of EBL absorption is negligible. {The lower panel of Fig. \ref{histo}  shows the distribution of predicted VHE $\gamma$-ray fluxes (> 200  GeV) of non-TeV BL Lacs together with different categories and  subclasses of known TeV BL Lacs. The brightest non-TeV BL Lacs occupy the region of known VHE emitters in their low state, making them potential targets for current IACTs.}

The \textit{VERITAS} telescope observed nine objects that are in our non-TeV candidate sample \citep{2016AJ....151..142A}. Four of these are in our most promising candidate list. The \textit{VERITAS} observations resulted only in upper limits and these upper limits are well above the minimum of $ S_{VHE,pred.}$. Comparing to median $ S_{VHE,pred.}$, we found three targets, TXS~0210+515, PKS~0829+046, and OJ~287, for which the reported flux upper limits were below our prediction. This is partly due to the scatter in our predictions derived from different correlations and partly due to the difficulty in comparing the two results because of the lack of spectral information in \citet{2016AJ....151..142A}, which leads to differences in the energy threshold between the two studies.

Since we originally formed our TeV sample, five more BL Lacs have been discovered in the VHE gamma rays. Firstly, 1ES~2037+521 was recently detected by \textit{MAGIC} telescopes \citep{2016ATel.9582....1M}. For this source, we calculated the median of predicted VHE $\gamma$-ray integral flux above 200 GeV to be $ (1.17 \pm 0.53) \times 10 ^{-11}$ [$\rm Ph/cm^2/s$]. The reported flux was $6 \times 10 ^{-12}$ [$\rm Ph/cm^2/s$], which is in a good agreement of our predicted value ($\sim 72\%$ probability of being the same). Moreover, four other objects, OJ~287, OT~081, S2~0109+22, and S4~0954+65 in the non-TeV sample were recently detected at VHE $\gamma$-rays during their high state \citep{2017ATel10051....1M,2016ATel.9267....1M,2015ATel.7844....1M,2015ATel.7080....1M}.  OJ~287 and S2~0109+22 are within the candidate list. The median predicted level of VHE $\gamma$-ray for OT~081 and S4~0954+65 are well above faintest VHE $\gamma$-ray flux and a bit lower than the sensitivity of current generation of IACTs. However, because of the lack of information we were not able to compare the observed flux with the predicted levels.

\section{Discussion and conclusions}
\label{sec5}

This paper presents the most extensive multi-wavelength data collection of TeV BL Lacs. We also present the first extensive correlation study between the low energy bands and the VHE $\gamma$-ray band using this extensive sample. We studied 35 correlations and found 8 significant ($p \le  1.5\times 10^{-3}$) correlations that include the VHE $\gamma$-ray band as the second component. Using these correlations we calculated the VHE $\gamma$-ray flux for 182 non-TeV BL Lacs. Finally, by sorting the non-TeV BL Lacs according to their predicted VHE $\gamma$-ray flux, we introduce a sample of promising TeV BL Lacs candidates.

\citet{2002A&A...384...56C} introduced 33 BL Lac candidates based on the similarity to the 5 known TeV BL Lacs in the radio and X-ray properties. They applied two different SSC models \citep{2002A&A...386..833G,1998MNRAS.299..433F} to calculate the level of TeV emission. \citet{2015A&A...579A..34A,2017A&A...598A..17C} (hereafter, 1WHSP and 2WHSP) introduced 76 and 136 promising high synchrotron peak (HSP) blazars by applying the figure of merit (FOM) criterion, which is defined as the ratio between the synchrotron peak flux ($\nu_{peak}F_{\nu_{peak}}$) of a source and the faintest TeV blazars. These sources were selected from a list of 992 and 1691 infrared colour-colour selected sources restricted by IR-radio and IR-X-ray flux ratios. \citet{2015MNRAS.446L..41P} applied a broadband SED Monte Carlo simulation method on the 1WHSP sample \citep{2015A&A...579A..34A} to select TeV BL Lac candidates. These authors assumed the synchrotron peak flux scales with the VHE flux. Applying typical sensitivity reachable by IACTs $\sim 14 $ mCrab., $\nu_{peak}>10^{15}$ [Hz], and EBL absorption \citep{2011MNRAS.410.2556D}, \citet{2015MNRAS.446L..41P} proposed 70 objects as TeV candidates. There are 28 HSP in our candidate list of which 10 are also introduced as good candidates ($FOM\geq 1$) in 1WHSP  and 2WHSP. This difference in the number of good candidates is due to  selection bias near the boundaries in 1WHSP and 2WHSP. For example, 1ES~2037+521, 4C~+42.22, and 1ES~0120+340 are HBLs based on their broadband SEDs. But their synchrotron peak frequency is affected by the contribution of host galaxy in the optical  band and therefore were not included in the 1WHSP and 2WHSP.   \citet{2014ApJS..215...14D} presented radio-loud candidates  $\gamma$-ray emitting blazars based on the measure of the  radio-to-IR flux ($ q22<-0.5$). In our non-TeV sample, 76 objects  have $ q22<-0.5$ while 21 of them are present in our candidate  list. \citet{2013ApJS..206...13M} selected HBLs for future TeV  observations (i.e. TBCs) based on the similarity of IR and X-ray  properties to the known TeV BL Lacs. The non-TeV sample shares 17 objects with TBCs sample of which 10 are present in the  candidate list. Finally, \citet{2011ApJ...739...73M} introduced 15  TeV BL Lacs candidates based on the similarity of curvature of their X-ray spectra to the TeV detected BL Lacs. There are 7 objects from this list in our non-TeV sample of which 4 are in our candidate list. 

In comparison to the above-mentioned methods, our empirical prediction method does not include major physical assumptions (e.g. location of synchrotron peak, lower energy bands brightness, and spectral properties in X-rays and/or $\gamma$-rays). Our method takes into account all of the important wavelengths for prediction and is not restricted to one specific wavelength. Therefore, we were able to introduce 21 BL Lac objects that were not proposed by any of the above-mentioned works. Moreover, the predicted VHE $\gamma$-ray flux are comparable with observations (5 confirmed VHE $\gamma$-ray detection) as discussed in section \ref{sec4}. Furthermore, the variability of the sources in VHE $\gamma$-ray band is taken into account using five different datasets in VHE $\gamma$-ray band. However, the simplistic method has known caveats:

\begin{itemize}
\item The lower energy data are not simultaneous to the VHE $\gamma$-ray. While we try to minimize the effect of this by considering different states of the VHE $\gamma$-ray emission, this does affect our correlation study, as discussed in detail in sections \ref{sigcor} and \ref{nonsig}.
 
\item Spectral indices in X-ray and $\gamma$-ray are not taken into account. Therefore, we might overestimate the VHE $\gamma$-ray flux for the sources with soft $\gamma$-ray and X-ray spectra.

\item The BL Lacs with unknown redshift are excluded from the non-TeV sample. Those with known redshift are likely to be at lower average redshift than those without redshift. Therefore, our search for TeV candidate is restricted to the relatively nearby BL Lacs.
 
\item VHE $\gamma$-ray fluxes of TeV sample are not corrected for the EBL absorption. The EBL corrected fluxes were not available in all publications and the EBL corrected fluxes reported in the literature had been corrected using a variety of EBL models, some of which were significantly outdated. While this is clearly incorrect physically and our VHE $\gamma$-ray sample extends up to redshift of $\sim$0.6, where the affect of the EBL absorption cannot be neglected, this was the only viable way to not to exclude a large number of sources from the study owing to missing information. As the majority of the sources are located at $z<0.3$, where the affect of the EBL absorption on the integral flux above 200\,GeV is relatively small, we expect that this does not bias our results significantly.
\end{itemize}

The method could be checked with the current generation of IACTs as we have discovered there are many candidates with the predicted VHE $\gamma$-ray flux level well above the IACTs sensitivity. It is notable that for the most promising candidates we assumed the minimum predicted VHE $\gamma$-ray level, which is the worst case assumption. Therefore the candidates should be detectable by the current generation of IACTs even in the low state. There are VHE $\gamma$-ray observations for a handful of our candidate VHE $\gamma$-ray sources. Within the 20 best candidates from our lists, there are upper limits reported by \textit{VERITAS} only for 5 sources \citep{2016AJ....151..142A}, which in all cases are above the fluxes we predicted. One of our candidate sources was recently detected by \textit{MAGIC} \citep{2016ATel.9582....1M}.

Follow-up observations are of interest for two main reasons. If the candidate sources are not detected with the current generation of IACTs even if deep observations are performed, it might indicate that TeV BL Lacs are a special class of BL Lacs objects. On the other hand, even if the population of TeV BL Lacs is finally large enough to perform a population study such as this work, it is still small enough for each new source to shed new light on the population and also to make prediction methods more accurate. Following up on these sources, which are most similar to known TeV BL Lacs, will also help to make the BL Lac sample more uniform. Finally, our weaker candidates provide a list that could be observed with next generation of IACTs, such as the Cherenkov Telescope Array\footnote{\url{www.cta-observatory.org}}.

\bibliographystyle{aa}
\bibliography{ref.bib}

\begin{thebibliography}{92}
\expandafter\ifx\csname natexlab\endcsname\relax\def\natexlab#1{#1}\fi

\bibitem[{Abdo {et~al.}(2010)Abdo, Ackermann, Agudo, Ajello, Aller, Aller,
  Angelakis, Arkharov, Axelsson, Bach, Baldini, Ballet, Barbiellini, Bastieri,
  Baughman, Bechtol, \& Bellazzini}]{0004-637X-716-1-30}
Abdo, A.~A., Ackermann, M., Agudo, I., {et~al.} 2010, \apj, 716, 30

\bibitem[{{Abdo} {et~al.}(2010){Abdo}, {Ackermann}, {Agudo}, {Ajello}, {Aller},
  {Aller}, {Angelakis}, {Arkharov}, {Axelsson}, {Bach}, \&
  et~al.}]{2010ApJ...716...30A}
{Abdo}, A.~A., {Ackermann}, M., {Agudo}, I., {et~al.} 2010, \apj, 716, 30

\bibitem[{{Acciari} {et~al.}(2011){Acciari}, {Aliu}, {Arlen}, {Aune},
  {Beilicke}, {Benbow}, {Boltuch}, {Bradbury}, {Buckley}, {Bugaev}, {Byrum},
  {Cannon}, {Cesarini}, {Ciupik}, {Cui}, {Dickherber}, {Duke}, {Falcone},
  {Finley}, {Finnegan}, {Fortson}, {Furniss}, {Galante}, {Gall}, {Gillanders},
  {Godambe}, {Grube}, {Guenette}, {Gyuk}, {Hanna}, {Holder}, {Hui}, {Humensky},
  {Imran}, {Kaaret}, {Karlsson}, {Kertzman}, {Kieda}, {Konopelko},
  {Krawczynski}, {Krennrich}, {Lang}, {Maier}, {McArthur}, {McCutcheon},
  {Moriarty}, {Ong}, {Otte}, {Ouellette}, {Pandel}, {Perkins}, {Pichel},
  {Pohl}, {Quinn}, {Ragan}, {Reyes}, {Reynolds}, {Roache}, {Rose}, {Rovero},
  {Schroedter}, {Sembroski}, {Senturk}, {Steele}, {Swordy}, {Theiling},
  {Thibadeau}, {Varlotta}, {Vassiliev}, {Vincent}, {Wagner}, {Wakely}, {Ward},
  {Weekes}, {Weinstein}, {Weisgarber}, {Williams}, {Wissel}, {Wood}, {Zitzer},
  {Garson}, {Lee}, {Sadun}, {Carini}, {Barnaby}, {Cook}, {Maune}, {Pease},
  {Smith}, {Walters}, {Berdyugin}, {Lindfors}, {Nilsson}, {Pasanen}, {Sainio},
  {Sillanpaa}, {Takalo}, {Villforth}, {Montaruli}, {Baker}, {Lahteenmaki},
  {Tornikoski}, {Hovatta}, {Nieppola}, {Aller}, \&
  {Aller}}]{2011ApJ...738...25A}
{Acciari}, V.~A., {Aliu}, E., {Arlen}, T., {et~al.} 2011, \apj, 738, 25

\bibitem[{{Acero} {et~al.}(2015){Acero}, {Ackermann}, {Ajello}, {Albert},
  {Atwood}, {Axelsson}, {Baldini}, {Ballet}, {Barbiellini}, {Bastieri},
  {Belfiore}, {Bellazzini}, {Bissaldi}, {Blandford}, {Bloom}, {Bogart},
  {Bonino}, {Bottacini}, {Bregeon}, {Britto}, {Bruel}, {Buehler}, {Burnett},
  {Buson}, {Caliandro}, {Cameron}, {Caputo}, {Caragiulo}, {Caraveo},
  {Casandjian}, {Cavazzuti}, {Charles}, {Chaves}, {Chekhtman}, {Cheung},
  {Chiang}, {Chiaro}, {Ciprini}, {Claus}, {Cohen-Tanugi}, {Cominsky}, {Conrad},
  {Cutini}, {D'Ammando}, {de Angelis}, {DeKlotz}, {de Palma}, {Desiante},
  {Digel}, {Di Venere}, {Drell}, {Dubois}, {Dumora}, {Favuzzi}, {Fegan},
  {Ferrara}, {Finke}, {Franckowiak}, {Fukazawa}, {Funk}, {Fusco}, {Gargano},
  {Gasparrini}, {Giebels}, {Giglietto}, {Giommi}, {Giordano}, {Giroletti},
  {Glanzman}, {Godfrey}, {Grenier}, {Grondin}, {Grove}, {Guillemot}, {Guiriec},
  {Hadasch}, {Harding}, {Hays}, {Hewitt}, {Hill}, {Horan}, {Iafrate}, {Jogler},
  {J{\'o}hannesson}, {Johnson}, {Johnson}, {Johnson}, {Johnson}, {Kamae},
  {Kataoka}, {Katsuta}, {Kuss}, {La Mura}, {Landriu}, {Larsson}, {Latronico},
  {Lemoine-Goumard}, {Li}, {Li}, {Longo}, {Loparco}, {Lott}, {Lovellette},
  {Lubrano}, {Madejski}, {Massaro}, {Mayer}, {Mazziotta}, {McEnery},
  {Michelson}, {Mirabal}, {Mizuno}, {Moiseev}, {Mongelli}, {Monzani},
  {Morselli}, {Moskalenko}, {Murgia}, {Nuss}, {Ohno}, {Ohsugi}, {Omodei},
  {Orienti}, {Orlando}, {Ormes}, {Paneque}, {Panetta}, {Perkins},
  {Pesce-Rollins}, {Piron}, {Pivato}, {Porter}, {Racusin}, {Rando}, {Razzano},
  {Razzaque}, {Reimer}, {Reimer}, {Reposeur}, {Rochester}, {Romani},
  {Salvetti}, {S{\'a}nchez-Conde}, {Saz Parkinson}, {Schulz}, {Siskind},
  {Smith}, {Spada}, {Spandre}, {Spinelli}, {Stephens}, {Strong}, {Suson},
  {Takahashi}, {Takahashi}, {Tanaka}, {Thayer}, {Thayer}, {Thompson},
  {Tibaldo}, {Tibolla}, {Torres}, {Torresi}, {Tosti}, {Troja}, {Van Klaveren},
  {Vianello}, {Winer}, {Wood}, {Wood}, {Zimmer}, \& {Fermi-LAT
  Collaboration}}]{2015ApJS..218...23A}
{Acero}, F., {Ackermann}, M., {Ajello}, M., {et~al.} 2015, \apjs, 218, 23

\bibitem[{{Ackermann} {et~al.}(2011){Ackermann}, {Ajello}, {Allafort},
  {Antolini}, {Atwood}, {Axelsson}, {Baldini}, {Ballet}, {Barbiellini},
  {Bastieri}, {Bechtol}, {Bellazzini}, {Berenji}, {Blandford}, {Bloom},
  {Bonamente}, {Borgland}, {Bottacini}, {Bouvier}, {Bregeon}, {Brigida},
  {Bruel}, {Buehler}, {Burnett}, {Buson}, {Caliandro}, {Cameron}, {Caraveo},
  {Casandjian}, {Cavazzuti}, {Cecchi}, {Charles}, {Cheung}, {Chiang},
  {Ciprini}, {Claus}, {Cohen-Tanugi}, {Conrad}, {Costamante}, {Cutini}, {de
  Angelis}, {de Palma}, {Dermer}, {Digel}, {Silva}, {Drell}, {Dubois},
  {Escande}, {Favuzzi}, {Fegan}, {Ferrara}, {Finke}, {Focke}, {Fortin},
  {Frailis}, {Fukazawa}, {Funk}, {Fusco}, {Gargano}, {Gasparrini}, {Gehrels},
  {Germani}, {Giebels}, {Giglietto}, {Giommi}, {Giordano}, {Giroletti},
  {Glanzman}, {Godfrey}, {Grenier}, {Grove}, {Guiriec}, {Gustafsson},
  {Hadasch}, {Hayashida}, {Hays}, {Healey}, {Horan}, {Hou}, {Hughes},
  {Iafrate}, {J{\'o}hannesson}, {Johnson}, {Johnson}, {Kamae}, {Katagiri},
  {Kataoka}, {Kn{\"o}dlseder}, {Kuss}, {Lande}, {Larsson}, {Latronico},
  {Longo}, {Loparco}, {Lott}, {Lovellette}, {Lubrano}, {Madejski}, {Mazziotta},
  {McConville}, {McEnery}, {Michelson}, {Mitthumsiri}, {Mizuno}, {Moiseev},
  {Monte}, {Monzani}, {Moretti}, {Morselli}, {Moskalenko}, {Murgia},
  {Nakamori}, {Naumann-Godo}, {Nolan}, {Norris}, {Nuss}, {Ohno}, {Ohsugi},
  {Okumura}, {Omodei}, {Orienti}, {Orlando}, {Ormes}, {Ozaki}, {Paneque},
  {Parent}, {Pesce-Rollins}, {Pierbattista}, {Piranomonte}, {Piron}, {Pivato},
  {Porter}, {Rain{\`o}}, {Rando}, {Razzano}, {Razzaque}, {Reimer}, {Reimer},
  {Ritz}, {Rochester}, {Romani}, {Roth}, {Sanchez}, {Sbarra}, {Scargle},
  {Schalk}, {Sgr{\`o}}, {Shaw}, {Siskind}, {Spandre}, {Spinelli}, {Strong},
  {Suson}, {Tajima}, {Takahashi}, {Takahashi}, {Tanaka}, {Thayer}, {Thayer},
  {Thompson}, {Tibaldo}, {Tinivella}, {Torres}, {Tosti}, {Troja}, {Uchiyama},
  {Vandenbroucke}, {Vasileiou}, {Vianello}, {Vitale}, {Waite}, {Wallace},
  {Wang}, {Winer}, {Wood}, {Wood}, \& {Zimmer}}]{2011ApJ...743..171A}
{Ackermann}, M., {Ajello}, M., {Allafort}, A., {et~al.} 2011, \apj, 743, 171

\bibitem[{{Ackermann} {et~al.}(2012){Ackermann}, {Ajello}, {Allafort},
  {Schady}, {Baldini}, {Ballet}, {Barbiellini}, {Bastieri}, {Bellazzini},
  {Blandford}, {Bloom}, {Borgland}, {Bottacini}, {Bouvier}, {Bregeon},
  {Brigida}, {Bruel}, {Buehler}, {Buson}, {Caliandro}, {Cameron}, {Caraveo},
  {Cavazzuti}, {Cecchi}, {Charles}, {Chaves}, {Chekhtman}, {Cheung}, {Chiang},
  {Chiaro}, {Ciprini}, {Claus}, {Cohen-Tanugi}, {Conrad}, {Cutini},
  {D'Ammando}, {de Palma}, {Dermer}, {Digel}, {do Couto e Silva},
  {Dom{\'{\i}}nguez}, {Drell}, {Drlica-Wagner}, {Favuzzi}, {Fegan}, {Focke},
  {Franckowiak}, {Fukazawa}, {Funk}, {Fusco}, {Gargano}, {Gasparrini},
  {Gehrels}, {Germani}, {Giglietto}, {Giordano}, {Giroletti}, {Glanzman},
  {Godfrey}, {Grenier}, {Grove}, {Guiriec}, {Gustafsson}, {Hadasch},
  {Hayashida}, {Hays}, {Jackson}, {Jogler}, {Kataoka}, {Kn{\"o}dlseder},
  {Kuss}, {Lande}, {Larsson}, {Latronico}, {Longo}, {Loparco}, {Lovellette},
  {Lubrano}, {Mazziotta}, {McEnery}, {Mehault}, {Michelson}, {Mizuno}, {Monte},
  {Monzani}, {Morselli}, {Moskalenko}, {Murgia}, {Tramacere}, {Nuss},
  {Greiner}, {Ohno}, {Ohsugi}, {Omodei}, {Orienti}, {Orlando}, {Ormes},
  {Paneque}, {Perkins}, {Pesce-Rollins}, {Piron}, {Pivato}, {Porter},
  {Rain{\`o}}, {Rando}, {Razzano}, {Razzaque}, {Reimer}, {Reimer}, {Reyes},
  {Ritz}, {Rau}, {Romoli}, {Roth}, {S{\'a}nchez-Conde}, {Sanchez}, {Scargle},
  {Sgr{\`o}}, {Siskind}, {Spandre}, {Spinelli}, {Stawarz}, {Suson},
  {Takahashi}, {Tanaka}, {Thayer}, {Thompson}, {Tibaldo}, {Tinivella},
  {Torres}, {Tosti}, {Troja}, {Usher}, {Vandenbroucke}, {Vasileiou},
  {Vianello}, {Vitale}, {Waite}, {Winer}, {Wood}, \&
  {Wood}}]{2012Sci...338.1190A}
{Ackermann}, M., {Ajello}, M., {Allafort}, A., {et~al.} 2012, Science, 338,
  1190

\bibitem[{{Aharonian} {et~al.}(2008){Aharonian}, {Akhperjanian}, {Barres de
  Almeida}, {Bazer-Bachi}, {Behera}, {Beilicke}, {Benbow}, {Bernl{\"o}hr},
  {Boisson}, {Borrel}, {Braun}, {Brion}, {Brucker}, {B{\"u}hler}, {Bulik},
  {B{\"u}sching}, {Boutelier}, {Carrigan}, {Chadwick}, {Chaves}, {Chounet},
  {Clapson}, {Coignet}, {Cornils}, {Costamante}, {Dalton}, {Degrange},
  {Dickinson}, {Djannati-Ata{\"i}}, {Domainko}, {O'C.~Drury}, {Dubois},
  {Dubus}, {Dyks}, {Egberts}, {Emmanoulopoulos}, {Espigat}, {Farnier},
  {Feinstein}, {Fiasson}, {F{\"o}rster}, {Fontaine}, {F{\"u}{\ss}ling},
  {Gabici}, {Gallant}, {Giebels}, {Glicenstein}, {Gl{\"u}ck}, {Goret},
  {Hadjichristidis}, {Hauser}, {Hauser}, {Heinzelmann}, {Henri}, {Hermann},
  {Hinton}, {Hoffmann}, {Hofmann}, {Holleran}, {Hoppe}, {Horns},
  {Jacholkowska}, {de Jager}, {Jung}, {Katarzy{\'n}ski}, {Kaufmann},
  {Kendziorra}, {Kerschhaggl}, {Khangulyan}, {Kh{\'e}lifi}, {Keogh}, {Komin},
  {Kosack}, {Lamanna}, {Latham}, {Lenain}, {Lohse}, {Martin},
  {Martineau-Huynh}, {Marcowith}, {Masterson}, {Maurin}, {McComb}, {Moderski},
  {Moulin}, {Naumann-Godo}, {de Naurois}, {Nedbal}, {Nekrassov}, {Nolan},
  {Ohm}, {Olive}, {de O{\~n}a Wilhelmi}, {Orford}, {Osborne}, {Ostrowski},
  {Panter}, {Pedaletti}, {Pelletier}, {Petrucci}, {Pita}, {P{\"u}hlhofer},
  {Punch}, {Quirrenbach}, {Raubenheimer}, {Raue}, {Rayner}, {Renaud}, {Rieger},
  {Ripken}, {Rob}, {Rosier-Lees}, {Rowell}, {Rudak}, {Ruppel}, {Sahakian},
  {Santangelo}, {Schlickeiser}, {Sch{\"o}ck}, {Schr{\"o}der}, {Schwanke},
  {Schwarzburg}, {Schwemmer}, {Shalchi}, {Sol}, {Spangler}, {Stawarz},
  {Steenkamp}, {Stegmann}, {Superina}, {Tam}, {Tavernet}, {Terrier}, {van
  Eldik}, {Vasileiadis}, {Venter}, {Vialle}, {Vincent}, {Vivier}, {V{\"o}lk},
  {Volpe}, {Wagner}, {Ward}, {Zdziarski}, \& {Zech}}]{2008A&A...481L.103A}
{Aharonian}, F., {Akhperjanian}, A.~G., {Barres de Almeida}, U., {et~al.} 2008,
  \aap, 481, L103

\bibitem[{{Aharonian}(2000)}]{2000NewA....5..377A}
{Aharonian}, F.~A. 2000, \na, 5, 377

\bibitem[{{Ahnen} {et~al.}(2016){Ahnen}, {Ansoldi}, {Antonelli}, {Antoranz},
  {Babic}, {Banerjee}, {Bangale}, {Barres de Almeida}, {Barrio}, {Becerra
  Gonz{\'a}lez}, {Bednarek}, {Bernardini}, {Biasuzzi}, {Biland}, {Blanch},
  {Bonnefoy}, {Bonnoli}, {Borracci}, {Bretz}, {Buson}, {Carosi}, {Chatterjee},
  {Clavero}, {Colin}, {Colombo}, {Contreras}, {Cortina}, {Covino}, {Da Vela},
  {Dazzi}, {De Angelis}, {De Lotto}, {de O{\~n}a Wilhelmi}, {Di Pierro},
  {Doert}, {Dom{\'{\i}}nguez}, {Dominis Prester}, {Dorner}, {Doro}, {Einecke},
  {Eisenacher Glawion}, {Elsaesser}, {Fallah Ramazani}, {Fern{\'a}ndez-Barral},
  {Fidalgo}, {Fonseca}, {Font}, {Frantzen}, {Fruck}, {Galindo}, {Garc{\'{\i}}a
  L{\'o}pez}, {Garczarczyk}, {Garrido Terrats}, {Gaug}, {Giammaria},
  {Godinovi{\'c}}, {Gonz{\'a}lez Mu{\~n}oz}, {Gora}, {Guberman}, {Hadasch},
  {Hahn}, {Hanabata}, {Hayashida}, {Herrera}, {Hose}, {Hrupec}, {Hughes},
  {Idec}, {Kodani}, {Konno}, {Kubo}, {Kushida}, {La Barbera}, {Lelas},
  {Lindfors}, {Lombardi}, {Longo}, {L{\'o}pez}, {L{\'o}pez-Coto}, {Majumdar},
  {Makariev}, {Mallot}, {Maneva}, {Manganaro}, {Mannheim}, {Maraschi},
  {Marcote}, {Mariotti}, {Mart{\'{\i}}nez}, {Mazin}, {Menzel}, {Miranda},
  {Mirzoyan}, {Moralejo}, {Moretti}, {Nakajima}, {Neustroev}, {Niedzwiecki},
  {Nievas Rosillo}, {Nilsson}, {Nishijima}, {Noda}, {Nogu{\'e}s}, {Orito},
  {Overkemping}, {Paiano}, {Palacio}, {Palatiello}, {Paneque}, {Paoletti},
  {Paredes}, {Paredes-Fortuny}, {Pedaletti}, {Perri}, {Persic}, {Poutanen},
  {Prada Moroni}, {Prandini}, {Puljak}, {Rhode}, {Rib{\'o}}, {Rico}, {Rodriguez
  Garcia}, {Saito}, {Satalecka}, {Schultz}, {Schweizer}, {Sillanp{\"a}{\"a}},
  {Sitarek}, {Snidaric}, {Sobczynska}, {Stamerra}, {Steinbring}, {Strzys},
  {Takalo}, {Takami}, {Tavecchio}, {Temnikov}, {Terzi{\'c}}, {Tescaro},
  {Teshima}, {Thaele}, {Torres}, {Toyama}, {Treves}, {Verguilov}, {Vovk},
  {Ward}, {Will}, {Wu}, {Zanin}, {D'Ammando}, {Berdyugin}, {Hovatta},
  {Max-Moerbeck}, {Raiteri}, {Readhead}, {Reinthal}, {Richards}, {Verrecchia},
  \& {Villata}}]{2016MNRAS.459.3271A}
{Ahnen}, M.~L., {Ansoldi}, S., {Antonelli}, L.~A., {et~al.} 2016, \mnras, 459,
  3271

\bibitem[{{Akritas} \& {Siebert}(1996)}]{1996MNRAS.278..919A}
{Akritas}, M.~G. \& {Siebert}, J. 1996, \mnras, 278, 919

\bibitem[{{Aleksi{\'c}} {et~al.}(2012){Aleksi{\'c}}, {Alvarez}, {Antonelli},
  {Antoranz}, {Ansoldi}, {Asensio}, {Backes}, {Barres de Almeida}, {Barrio},
  {Bastieri}, {Becerra Gonz{\'a}lez}, {Bednarek}, {Berger}, {Bernardini},
  {Biland}, {Blanch}, {Bock}, {Boller}, {Bonnoli}, {Borla Tridon}, {Bretz},
  {Ca{\~n}ellas}, {Carmona}, {Carosi}, {Colin}, {Colombo}, {Contreras},
  {Cortina}, {Cossio}, {Covino}, {Da Vela}, {Dazzi}, {De Angelis}, {De Caneva},
  {De Cea del Pozo}, {De Lotto}, {Delgado Mendez}, {Diago Ortega}, {Doert},
  {Dom{\'{\i}}nguez}, {Dominis Prester}, {Dorner}, {Doro}, {Eisenacher},
  {Elsaesser}, {Ferenc}, {Fonseca}, {Font}, {Fruck}, {Garc{\'{\i}}a L{\'o}pez},
  {Garczarczyk}, {Garrido Terrats}, {Gaug}, {Giavitto}, {Godinovi{\'c}},
  {Gonz{\'a}lez Mu{\~n}oz}, {Gozzini}, {Hadasch}, {H{\"a}fner}, {Herrero},
  {Hildebrand}, {Hose}, {Hrupec}, {Huber}, {Jankowski}, {Jogler}, {Kadenius},
  {Kellermann}, {Klepser}, {Kr{\"a}henb{\"u}hl}, {Krause}, {La Barbera},
  {Lelas}, {Leonardo}, {Lewandowska}, {Lindfors}, {Lombardi}, {L{\'o}pez},
  {L{\'o}pez-Coto}, {L{\'o}pez-Oramas}, {Lorenz}, {Makariev}, {Maneva},
  {Mankuzhiyil}, {Mannheim}, {Maraschi}, {Mariotti}, {Mart{\'{\i}}nez},
  {Mazin}, {Meucci}, {Miranda}, {Mirzoyan}, {Mold{\'o}n}, {Moralejo},
  {Munar-Adrover}, {Niedzwiecki}, {Nieto}, {Nilsson}, {Nowak}, {Orito},
  {Paiano}, {Paneque}, {Paoletti}, {Pardo}, {Paredes}, {Partini},
  {Perez-Torres}, {Persic}, {Pilia}, {Pochon}, {Prada}, {Prada Moroni},
  {Prandini}, {Puerto Gimenez}, {Puljak}, {Reichardt}, {Reinthal}, {Rhode},
  {Rib{\'o}}, {Rico}, {R{\"u}gamer}, {Saggion}, {Saito}, {Saito}, {Salvati},
  {Satalecka}, {Scalzotto}, {Scapin}, {Schultz}, {Schweizer}, {Shore},
  {Sillanp{\"a}{\"a}}, {Sitarek}, {Snidaric}, {Sobczynska}, {Spanier}, {Spiro},
  {Stamatescu}, {Stamerra}, {Steinke}, {Storz}, {Strah}, {Sun}, {Suri{\'c}},
  {Takalo}, {Takami}, {Tavecchio}, {Temnikov}, {Terzi{\'c}}, {Tescaro},
  {Teshima}, {Tibolla}, {Torres}, {Treves}, {Uellenbeck}, {Vogler}, {Wagner},
  {Weitzel}, {Zabalza}, {Zandanel}, {Zanin}, {Berdyugin}, {Buson},
  {J{\"a}rvel{\"a}}, {Larsson}, {L{\"a}hteenm{\"a}ki}, \&
  {Tammi}}]{2012A&A...544A.142A}
{Aleksi{\'c}}, J., {Alvarez}, E.~A., {Antonelli}, L.~A., {et~al.} 2012, \aap,
  544, A142

\bibitem[{{Aleksi{\'c}} {et~al.}(2016){Aleksi{\'c}}, {Ansoldi}, {Antonelli},
  {Antoranz}, {Babic}, {Bangale}, {Barcel{\'o}}, {Barrio}, {Becerra
  Gonz{\'a}lez}, {Bednarek}, {Bernardini}, {Biasuzzi}, {Biland}, {Bitossi},
  {Blanch}, {Bonnefoy}, {Bonnoli}, {Borracci}, {Bretz}, {Carmona}, {Carosi},
  {Cecchi}, {Colin}, {Colombo}, {Contreras}, {Corti}, {Cortina}, {Covino}, {Da
  Vela}, {Dazzi}, {De Angelis}, {De Caneva}, {De Lotto}, {de O{\~n}a Wilhelmi},
  {Delgado Mendez}, {Dettlaff}, {Dominis Prester}, {Dorner}, {Doro}, {Einecke},
  {Eisenacher}, {Elsaesser}, {Fidalgo}, {Fink}, {Fonseca}, {Font}, {Frantzen},
  {Fruck}, {Galindo}, {Garc{\'{\i}}a L{\'o}pez}, {Garczarczyk}, {Garrido
  Terrats}, {Gaug}, {Giavitto}, {Godinovi{\'c}}, {Gonz{\'a}lez Mu{\~n}oz},
  {Gozzini}, {Haberer}, {Hadasch}, {Hanabata}, {Hayashida}, {Herrera},
  {Hildebrand}, {Hose}, {Hrupec}, {Idec}, {Illa}, {Kadenius}, {Kellermann},
  {Knoetig}, {Kodani}, {Konno}, {Krause}, {Kubo}, {Kushida}, {La Barbera},
  {Lelas}, {Lemus}, {Lewandowska}, {Lindfors}, {Lombardi}, {Longo},
  {L{\'o}pez}, {L{\'o}pez-Coto}, {L{\'o}pez-Oramas}, {Lorca}, {Lorenz},
  {Lozano}, {Makariev}, {Mallot}, {Maneva}, {Mankuzhiyil}, {Mannheim},
  {Maraschi}, {Marcote}, {Mariotti}, {Mart{\'{\i}}nez}, {Mazin}, {Menzel},
  {Miranda}, {Mirzoyan}, {Moralejo}, {Munar-Adrover}, {Nakajima}, {Negrello},
  {Neustroev}, {Niedzwiecki}, {Nilsson}, {Nishijima}, {Noda}, {Orito},
  {Overkemping}, {Paiano}, {Palatiello}, {Paneque}, {Paoletti}, {Paredes},
  {Paredes-Fortuny}, {Persic}, {Poutanen}, {Prada Moroni}, {Prandini},
  {Puljak}, {Reinthal}, {Rhode}, {Rib{\'o}}, {Rico}, {Rodriguez Garcia},
  {R{\"u}gamer}, {Saito}, {Saito}, {Satalecka}, {Scalzotto}, {Scapin},
  {Schultz}, {Schlammer}, {Schmidl}, {Schweizer}, {Shore}, {Sillanp{\"a}{\"a}},
  {Sitarek}, {Snidaric}, {Sobczynska}, {Spanier}, {Stamerra}, {Steinbring},
  {Storz}, {Strzys}, {Takalo}, {Takami}, {Tavecchio}, {Tejedor}, {Temnikov},
  {Terzi{\'c}}, {Tescaro}, {Teshima}, {Thaele}, {Tibolla}, {Torres}, {Toyama},
  {Treves}, {Vogler}, {Wetteskind}, {Will}, \& {Zanin}}]{2016APh....72...76A}
{Aleksi{\'c}}, J., {Ansoldi}, S., {Antonelli}, L.~A., {et~al.} 2016,
  Astroparticle Physics, 72, 76

\bibitem[{{Aleksi{\'c}} {et~al.}(2014){Aleksi{\'c}}, {Ansoldi}, {Antonelli},
  {Antoranz}, {Babic}, {Bangale}, {Barres de Almeida}, {Barrio}, {Becerra
  Gonz{\'a}lez}, {Bednarek}, {Berger}, {Bernardini}, {Biland}, {Blanch},
  {Bock}, {Bonnefoy}, {Bonnoli}, {Borracci}, {Bretz}, {Carmona}, {Carosi},
  {Carreto Fidalgo}, {Colin}, {Colombo}, {Contreras}, {Cortina}, {Covino}, {Da
  Vela}, {Dazzi}, {De Angelis}, {De Caneva}, {De Lotto}, {Delgado Mendez},
  {Doert}, {Dom{\'{\i}}nguez}, {Dominis Prester}, {Dorner}, {Doro}, {Einecke},
  {Eisenacher}, {Elsaesser}, {Farina}, {Ferenc}, {Fonseca}, {Font}, {Frantzen},
  {Fruck}, {Garc{\'{\i}}a L{\'o}pez}, {Garczarczyk}, {Garrido Terrats}, {Gaug},
  {Giavitto}, {Godinovi{\'c}}, {Gonz{\'a}lez Mu{\~n}oz}, {Gozzini}, {Hadasch},
  {Hayashida}, {Herrero}, {Hildebrand}, {Hose}, {Hrupec}, {Idec}, {Kadenius},
  {Kellermann}, {Kodani}, {Konno}, {Krause}, {Kubo}, {Kushida}, {La Barbera},
  {Lelas}, {Lewandowska}, {Lindfors}, {Lombardi}, {L{\'o}pez},
  {L{\'o}pez-Coto}, {L{\'o}pez-Oramas}, {Lorenz}, {Lozano}, {Makariev},
  {Mallot}, {Maneva}, {Mankuzhiyil}, {Mannheim}, {Maraschi}, {Marcote},
  {Mariotti}, {Mart{\'{\i}}nez}, {Mazin}, {Menzel}, {Meucci}, {Miranda},
  {Mirzoyan}, {Moralejo}, {Munar-Adrover}, {Nakajima}, {Niedzwiecki},
  {Nilsson}, {Nishijima}, {Nowak}, {Orito}, {Overkemping}, {Paiano},
  {Palatiello}, {Paneque}, {Paoletti}, {Paredes}, {Paredes-Fortuny}, {Partini},
  {Persic}, {Prada}, {Prada Moroni}, {Prandini}, {Preziuso}, {Puljak},
  {Reinthal}, {Rhode}, {Rib{\'o}}, {Rico}, {Rodriguez Garcia}, {R{\"u}gamer},
  {Saggion}, {Saito}, {Saito}, {Salvati}, {Satalecka}, {Scalzotto}, {Scapin},
  {Schultz}, {Schweizer}, {Shore}, {Sillanp{\"a}{\"a}}, {Sitarek}, {Snidaric},
  {Sobczynska}, {Spanier}, {Stamatescu}, {Stamerra}, {Steinbring}, {Storz},
  {Sun}, {Suri{\'c}}, {Takalo}, {Takami}, {Tavecchio}, {Temnikov},
  {Terzi{\'c}}, {Tescaro}, {Teshima}, {Thaele}, {Tibolla}, {Torres}, {Toyama},
  {Treves}, {Uellenbeck}, {Vogler}, {Wagner}, {Zandanel}, {Zanin}, {MAGIC
  Collaboration}, {Cutini}, {Gasparrini}, {Furniss}, {Hovatta}, {Kangas},
  {Kankare}, {Kotilainen}, {Lister}, {L{\"a}hteenm{\"a}ki}, {Max-Moerbeck},
  {Pavlidou}, {Readhead}, \& {Richards}}]{2014A&A...567A.135A}
{Aleksi{\'c}}, J., {Ansoldi}, S., {Antonelli}, L.~A., {et~al.} 2014, \aap, 567,
  A135

\bibitem[{{Aleksi{\'c}} {et~al.}(2015{\natexlab{a}}){Aleksi{\'c}}, {Ansoldi},
  {Antonelli}, {Antoranz}, {Babic}, {Bangale}, {Barres de Almeida}, {Barrio},
  {Becerra Gonz{\'a}lez}, {Bednarek}, {Berger}, {Bernardini}, {Biland},
  {Blanch}, {Bonnefoy}, {Bonnoli}, {Borracci}, {Bretz}, {Carmona}, {Carosi},
  {Carreto Fidalgo}, {Colin}, {Colombo}, {Contreras}, {Cortina}, {Covino}, {da
  Vela}, {Dazzi}, {de Angelis}, {de Caneva}, {de Lotto}, {Delgado Mendez},
  {Doert}, {Dom{\'{\i}}nguez}, {Dominis Prester}, {Dorner}, {Doro}, {Einecke},
  {Eisenacher}, {Elsaesser}, {Farina}, {Ferenc}, {Fonseca}, {Font}, {Frantzen},
  {Fruck}, {Garc{\'{\i}}a L{\'o}pez}, {Garczarczyk}, {Garrido Terrats}, {Gaug},
  {Godinovi{\'c}}, {Gonz{\'a}lez Mu{\~n}oz}, {Gozzini}, {Hadasch}, {Hayashida},
  {Herrera}, {Herrero}, {Hildebrand}, {Hose}, {Hrupec}, {Idec}, {Kadenius},
  {Kellermann}, {Kodani}, {Konno}, {Krause}, {Kubo}, {Kushida}, {La Barbera},
  {Lelas}, {Lewandowska}, {Lindfors}, {Lombardi}, {L{\'o}pez},
  {L{\'o}pez-Coto}, {L{\'o}pez-Oramas}, {Lorenz}, {Lozano}, {Makariev},
  {Mallot}, {Maneva}, {Mankuzhiyil}, {Mannheim}, {Maraschi}, {Marcote},
  {Mariotti}, {Mart{\'{\i}}nez}, {Mazin}, {Menzel}, {Meucci}, {Miranda},
  {Mirzoyan}, {Moralejo}, {Munar-Adrover}, {Nakajima}, {Niedzwiecki},
  {Nilsson}, {Nishijima}, {Noda}, {Nowak}, {Orito}, {Overkemping}, {Paiano},
  {Palatiello}, {Paneque}, {Paoletti}, {Paredes}, {Paredes-Fortuny}, {Partini},
  {Persic}, {Prada}, {Moroni}, {Prandini}, {Preziuso}, {Puljak}, {Reinthal},
  {Rhode}, {Rib{\'o}}, {Rico}, {Rodriguez Garcia}, {R{\"u}gamer}, {Saggion},
  {Saito}, {Saito}, {Satalecka}, {Scalzotto}, {Scapin}, {Schultz}, {Schweizer},
  {Shore}, {Sillanp{\"a}{\"a}}, {Sitarek}, {Snidaric}, {Sobczynska}, {Spanier},
  {Stamatescu}, {Stamerra}, {Steinbring}, {Storz}, {Sun}, {Suri{\'c}},
  {Takalo}, {Takami}, {Tavecchio}, {Temnikov}, {Terzi{\'c}}, {Tescaro},
  {Teshima}, {Thaele}, {Tibolla}, {Torres}, {Toyama}, {Treves}, {Uellenbeck},
  {Vogler}, {Wagner}, {Zandanel}, {Zanin}, {MAGIC Collaboration}, {Tronconi},
  {Buson}, \& {Borghese}}]{2015MNRAS.446..217A}
{Aleksi{\'c}}, J., {Ansoldi}, S., {Antonelli}, L.~A., {et~al.}
  2015{\natexlab{a}}, \mnras, 446, 217

\bibitem[{{Aleksi{\'c}} {et~al.}(2015{\natexlab{b}}){Aleksi{\'c}}, {Ansoldi},
  {Antonelli}, {Antoranz}, {Babic}, {Bangale}, {Barres de Almeida}, {Barrio},
  {Becerra Gonz{\'a}lez}, {Bednarek}, \& et~al.}]{2015A&A...576A.126A}
{Aleksi{\'c}}, J., {Ansoldi}, S., {Antonelli}, L.~A., {et~al.}
  2015{\natexlab{b}}, \aap, 576, A126

\bibitem[{{Aleksi{\'c}} {et~al.}(2015{\natexlab{c}}){Aleksi{\'c}}, {Ansoldi},
  {Antonelli}, {Antoranz}, {Babic}, {Bangale}, {Barrio}, {Becerra
  Gonz{\'a}lez}, {Bednarek}, {Bernardini}, {Biasuzzi}, {Biland}, {Blanch},
  {Bonnefoy}, {Bonnoli}, {Borracci}, {Bretz}, {Carmona}, {Carosi}, {Colin},
  {Colombo}, {Contreras}, {Cortina}, {Covino}, {Da Vela}, {Dazzi}, {De
  Angelis}, {De Caneva}, {De Lotto}, {de O{\~n}a Wilhelmi}, {Delgado Mendez},
  {Di Pierro}, {Dominis Prester}, {Dorner}, {Doro}, {Einecke}, {Eisenacher},
  {Elsaesser}, {Fern{\'a}ndez-Barral}, {Fidalgo}, {Fonseca}, {Font},
  {Frantzen}, {Fruck}, {Galindo}, {Garc{\'{\i}}a L{\'o}pez}, {Garczarczyk},
  {Garrido Terrats}, {Gaug}, {Godinovi{\'c}}, {Gonz{\'a}lez Mu{\~n}oz},
  {Gozzini}, {Hadasch}, {Hanabata}, {Hayashida}, {Herrera}, {Hose}, {Hrupec},
  {Idec}, {Kadenius}, {Kellermann}, {Knoetig}, {Kodani}, {Konno}, {Krause},
  {Kubo}, {Kushida}, {La Barbera}, {Lelas}, {Lewandowska}, {Lindfors},
  {Lombardi}, {Longo}, {L{\'o}pez}, {L{\'o}pez-Coto}, {L{\'o}pez-Oramas},
  {Lorenz}, {Lozano}, {Makariev}, {Mallot}, {Maneva}, {Mannheim}, {Maraschi},
  {Marcote}, {Mariotti}, {Mart{\'{\i}}nez}, {Mazin}, {Menzel}, {Miranda},
  {Mirzoyan}, {Moralejo}, {Munar-Adrover}, {Nakajima}, {Neustroev},
  {Niedzwiecki}, {Nievas Rosillo}, {Nilsson}, {Nishijima}, {Noda}, {Orito},
  {Overkemping}, {Paiano}, {Palatiello}, {Paneque}, {Paoletti}, {Paredes},
  {Paredes-Fortuny}, {Persic}, {Poutanen}, {Prada Moroni}, {Prandini},
  {Puljak}, {Reinthal}, {Rhode}, {Rib{\'o}}, {Rico}, {Rodriguez Garcia},
  {Saito}, {Saito}, {Satalecka}, {Scalzotto}, {Scapin}, {Schultz}, {Schweizer},
  {Shore}, {Sillanp{\"a}{\"a}}, {Sitarek}, {Snidaric}, {Sobczynska},
  {Stamerra}, {Steinbring}, {Strzys}, {Takalo}, {Takami}, {Tavecchio},
  {Temnikov}, {Terzi{\'c}}, {Tescaro}, {Teshima}, {Thaele}, {Torres}, {Toyama},
  {Treves}, {Vogler}, {Will}, {Zanin}, {Berger}, {Buson}, {D'Ammando},
  {Gasparrini}, {Hovatta}, {Max-Moerbeck}, {Readhead}, \&
  {Richards}}]{2015MNRAS.451..739A}
{Aleksi{\'c}}, J., {Ansoldi}, S., {Antonelli}, L.~A., {et~al.}
  2015{\natexlab{c}}, \mnras, 451, 739

\bibitem[{{Aleksi{\'c}} {et~al.}(2011){Aleksi{\'c}}, {Antonelli}, {Antoranz},
  {Backes}, {Barrio}, {Bastieri}, {Becerra Gonz{\'a}lez}, {Bednarek},
  {Berdyugin}, {Berger}, {Bernardini}, {Biland}, {Blanch}, {Bock}, {Boller},
  {Bonnoli}, {Bordas}, {Borla Tridon}, {Bosch-Ramon}, {Bose}, {Braun}, {Bretz},
  {Camara}, {Ca{\~n}ellas}, {Carmona}, {Carosi}, {Colin}, {Colombo},
  {Contreras}, {Cortina}, {Cossio}, {Covino}, {Dazzi}, {De Angelis}, {De Cea
  del Pozo}, {De Lotto}, {De Maria}, {De Sabata}, {Delgado Mendez}, {Diago
  Ortega}, {Doert}, {Dom{\'{\i}}nguez}, {Dominis Prester}, {Dorner}, {Doro},
  {Elsaesser}, {Errando}, {Ferenc}, {Fonseca}, {Font}, {Garc{\'{\i}}a
  L{\'o}pez}, {Garczarczyk}, {Giavitto}, {Godinovi{\'c}}, {Hadasch}, {Herrero},
  {Hildebrand}, {H{\"o}hne-M{\"o}nch}, {Hose}, {Hrupec}, {Jogler}, {Klepser},
  {Kr{\"a}henb{\"u}hl}, {Kranich}, {Krause}, {La Barbera}, {Leonardo},
  {Lindfors}, {Lombardi}, {Longo}, {L{\'o}pez}, {Lorenz}, {Majumdar},
  {Makariev}, {Maneva}, {Mankuzhiyil}, {Mannheim}, {Maraschi}, {Mariotti},
  {Mart{\'{\i}}nez}, {Mazin}, {Meucci}, {Miranda}, {Mirzoyan}, {Miyamoto},
  {Mold{\'o}n}, {Moralejo}, {Nieto}, {Nilsson}, {Orito}, {Oya}, {Paoletti},
  {Paredes}, {Partini}, {Pasanen}, {Pauss}, {Pegna}, {Perez-Torres}, {Persic},
  {Peruzzo}, {Pochon}, {Prada}, {Prada Moroni}, {Prandini}, {Puchades},
  {Puljak}, {Reichardt}, {Reinthal}, {Rhode}, {Rib{\'o}}, {Rico},
  {R{\"u}gamer}, {Saggion}, {Saito}, {Saito}, {Salvati}, {S{\'a}nchez-Conde},
  {Satalecka}, {Scalzotto}, {Scapin}, {Schultz}, {Schweizer}, {Shayduk},
  {Shore}, {Sierpowska-Bartosik}, {Sillanp{\"a}{\"a}}, {Sitarek}, {Sobczynska},
  {Spanier}, {Spiro}, {Stamerra}, {Steinke}, {Storz}, {Strah}, {Struebig},
  {Suric}, {Takalo}, {Tavecchio}, {Temnikov}, {Terzi{\'c}}, {Tescaro},
  {Teshima}, {Thom}, {Torres}, {Vankov}, {Wagner}, {Weitzel}, {Zabalza},
  {Zandanel}, \& {Zanin}}]{2011ApJ...726...58A}
{Aleksi{\'c}}, J., {Antonelli}, L.~A., {Antoranz}, P., {et~al.} 2011, \apj,
  726, 58

\bibitem[{{Aliu} {et~al.}(2014){Aliu}, {Archambault}, {Arlen}, {Aune},
  {Barnacka}, {Beilicke}, {Benbow}, {Berger}, {Bird}, {Bouvier}, {Buckley},
  {Bugaev}, {Cerruti}, {Chen}, {Ciupik}, {Collins-Hughes}, {Connolly}, {Cui},
  {Dumm}, {Eisch}, {Falcone}, {Federici}, {Feng}, {Finley}, {Fleischhack},
  {Fortin}, {Fortson}, {Furniss}, {Galante}, {Gillanders}, {Griffin},
  {Griffiths}, {Grube}, {Gyuk}, {H{\aa}kansson}, {Hanna}, {Holder}, {Hughes},
  {Hughes}, {Humensky}, {Johnson}, {Kaaret}, {Kar}, {Kertzman}, {Khassen},
  {Kieda}, {Krawczynski}, {Krennrich}, {Lang}, {Madhavan}, {Majumdar},
  {McArthur}, {McCann}, {Meagher}, {Millis}, {Moriarty}, {Mukherjee}, {Nelson},
  {Nieto}, {O'Faol{\'a}in de Bhr{\'o}ithe}, {Ong}, {Otte}, {Park}, {Perkins},
  {Pohl}, {Popkow}, {Prokoph}, {Quinn}, {Ragan}, {Rajotte}, {Reyes},
  {Reynolds}, {Richards}, {Roache}, {Sadun}, {Santander}, {Sembroski},
  {Shahinyan}, {Sheidaei}, {Smith}, {Staszak}, {Telezhinsky}, {Theiling},
  {Tyler}, {Varlotta}, {Vassiliev}, {Vincent}, {Wakely}, {Weekes}, {Weinstein},
  {Welsing}, {Wilhelm}, {Williams}, {Zitzer}, {VERITAS Collaboration},
  {B{\"o}ttcher}, \& {Fumagalli}}]{2014ApJ...797...89A}
{Aliu}, E., {Archambault}, S., {Arlen}, T., {et~al.} 2014, \apj, 797, 89

\bibitem[{{Archambault} {et~al.}(2016){Archambault}, {Archer}, {Benbow},
  {Bird}, {Biteau}, {Buchovecky}, {Buckley}, {Bugaev}, {Byrum}, {Cerruti},
  {Chen}, {Ciupik}, {Connolly}, {Cui}, {Eisch}, {Errando}, {Falcone}, {Feng},
  {Finley}, {Fleischhack}, {Fortin}, {Fortson}, {Furniss}, {Gillanders},
  {Griffin}, {Grube}, {Gyuk}, {H{\"u}tten}, {H{\aa}kansson}, {Hanna}, {Holder},
  {Humensky}, {Johnson}, {Kaaret}, {Kar}, {Kelley-Hoskins}, {Kertzman},
  {Kieda}, {Krause}, {Krennrich}, {Kumar}, {Lang}, {Maier}, {McArthur},
  {McCann}, {Meagher}, {Moriarty}, {Mukherjee}, {Nguyen}, {Nieto},
  {O'Faol{\'a}in de Bhr{\'o}ithe}, {Ong}, {Otte}, {Park}, {Perkins}, {Pichel},
  {Pohl}, {Popkow}, {Pueschel}, {Quinn}, {Ragan}, {Reynolds}, {Richards},
  {Roache}, {Rovero}, {Santander}, {Sembroski}, {Shahinyan}, {Smith},
  {Staszak}, {Telezhinsky}, {Tucci}, {Tyler}, {Vincent}, {Wakely}, {Weiner},
  {Weinstein}, {Williams}, {Zitzer}, {VERITAS Collaboration}, {Fumagalli}, \&
  {Prochaska}}]{2016AJ....151..142A}
{Archambault}, S., {Archer}, A., {Benbow}, W., {et~al.} 2016, \aj, 151, 142

\bibitem[{{Arsioli} {et~al.}(2015){Arsioli}, {Fraga}, {Giommi}, {Padovani}, \&
  {Marrese}}]{2015A&A...579A..34A}
{Arsioli}, B., {Fraga}, B., {Giommi}, P., {Padovani}, P., \& {Marrese}, P.~M.
  2015, \aap, 579, A34

\bibitem[{{Becherini} {et~al.}(2012){Becherini}, {Boisson}, {Cerruti}, \&
  {H.E.S.S.~Collaboration}}]{10.1063/1.4772304}
{Becherini}, Y., {Boisson}, C., {Cerruti}, M., \& {H.E.S.S.~Collaboration}.
  2012, AIP Conference Proceedings, 1505, 490

\bibitem[{{Blandford} \& {K{\"o}nigl}(1979)}]{1979ApJ...232...34B}
{Blandford}, R.~D. \& {K{\"o}nigl}, A. 1979, \apj, 232, 34

\bibitem[{{Blandford} \& {Rees}(1978)}]{1978PhyS...17..265B}
{Blandford}, R.~D. \& {Rees}, M.~J. 1978, \physscr, 17, 265

\bibitem[{{Burrows} {et~al.}(2004){Burrows}, {Hill}, {Nousek}, {Wells},
  {Chincarini}, {Abbey}, {Beardmore}, {Bosworth}, {Br{\"a}uninger}, {Burkert},
  {Campana}, {Capalbi}, {Chang}, {Citterio}, {Freyberg}, {Giommi}, {Hartner},
  {Killough}, {Kittle}, {Klar}, {Mangels}, {McMeekin}, {Miles}, {Moretti},
  {Mori}, {Morris}, {Mukerjee}, {Osborne}, {Short}, {Tagliaferri},
  {Tamburelli}, {Watson}, {Willingale}, \& {Zugger}}]{2004SPIE.5165..201B}
{Burrows}, D.~N., {Hill}, J.~E., {Nousek}, J.~A., {et~al.} 2004, in \procspie,
  Vol. 5165, X-Ray and Gamma-Ray Instrumentation for Astronomy XIII, ed. K.~A.
  {Flanagan} \& O.~H.~W. {Siegmund}, 201--216

\bibitem[{{Chang} {et~al.}(2017){Chang}, {Arsioli}, {Giommi}, \&
  {Padovani}}]{2017A&A...598A..17C}
{Chang}, Y.-L., {Arsioli}, B., {Giommi}, P., \& {Padovani}, P. 2017, \aap, 598,
  A17

\bibitem[{{Cologna} {et~al.}(2015){Cologna}, {Mohamed}, {Wagner},
  {Wierzcholska}, {Romoli}, \& {Kurtanidze}}]{2015ICRC...34..762C}
{Cologna}, G., {Mohamed}, M., {Wagner}, S., {et~al.} 2015, in International
  Cosmic Ray Conference, Vol.~34, 34th International Cosmic Ray Conference
  (ICRC2015), 762

\bibitem[{{Coppi}(1992)}]{1992MNRAS.258..657C}
{Coppi}, P.~S. 1992, \mnras, 258, 657

\bibitem[{{Costamante} \& {Ghisellini}(2002)}]{2002A&A...384...56C}
{Costamante}, L. \& {Ghisellini}, G. 2002, \aap, 384, 56

\bibitem[{{D'Abrusco} {et~al.}(2012){D'Abrusco}, {Massaro}, {Ajello},
  {Grindlay}, {Smith}, \& {Tosti}}]{2012ApJ...748...68D}
{D'Abrusco}, R., {Massaro}, F., {Ajello}, M., {et~al.} 2012, \apj, 748, 68

\bibitem[{{D'Abrusco} {et~al.}(2014){D'Abrusco}, {Massaro}, {Paggi}, {Smith},
  {Masetti}, {Landoni}, \& {Tosti}}]{2014ApJS..215...14D}
{D'Abrusco}, R., {Massaro}, F., {Paggi}, A., {et~al.} 2014, \apjs, 215, 14

\bibitem[{{D'Elia} {et~al.}(2013){D'Elia}, {Perri}, {Puccetti}, {Capalbi},
  {Giommi}, {Burrows}, {Campana}, {Tagliaferri}, {Cusumano}, {Evans},
  {Gehrels}, {Kennea}, {Moretti}, {Nousek}, {Osborne}, {Romano}, \&
  {Stratta}}]{2013A&A...551A.142D}
{D'Elia}, V., {Perri}, M., {Puccetti}, S., {et~al.} 2013, \aap, 551, A142

\bibitem[{{Dermer} \& {Schlickeiser}(1994)}]{1994ApJS...90..945D}
{Dermer}, C.~D. \& {Schlickeiser}, R. 1994, \apjs, 90, 945

\bibitem[{{Djannati-Ata{\"i}}(2009)}]{2009NIMPA.602...28D}
{Djannati-Ata{\"i}}, A. 2009, Nuclear Instruments and Methods in Physics
  Research A, 602, 28

\bibitem[{{Dom{\'{\i}}nguez} {et~al.}(2011){Dom{\'{\i}}nguez}, {Primack},
  {Rosario}, {Prada}, {Gilmore}, {Faber}, {Koo}, {Somerville},
  {P{\'e}rez-Torres}, {P{\'e}rez-Gonz{\'a}lez}, {Huang}, {Davis},
  {Guhathakurta}, {Barmby}, {Conselice}, {Lozano}, {Newman}, \&
  {Cooper}}]{2011MNRAS.410.2556D}
{Dom{\'{\i}}nguez}, A., {Primack}, J.~R., {Rosario}, D.~J., {et~al.} 2011,
  \mnras, 410, 2556

\bibitem[{{Donato} {et~al.}(2005){Donato}, {Sambruna}, \&
  {Gliozzi}}]{2005A&A...433.1163D}
{Donato}, D., {Sambruna}, R.~M., \& {Gliozzi}, M. 2005, \aap, 433, 1163

\bibitem[{Efron \& Stein(1981)}]{tEFR81a}
Efron, B. \& Stein, C. 1981, Annals of Statistics, 9, 586

\bibitem[{Efron \& Tibshirani(1993)}]{tEFR93a}
Efron, B. \& Tibshirani, R.~J. 1993, An Introduction to the Bootstrap (New
  York, NY: Chapman \& Hall)

\bibitem[{{Fan} {et~al.}(2012){Fan}, {Bai}, {Liu}, {Chen}, \&
  {Liao}}]{2012RAA....12.1475F}
{Fan}, X.-L., {Bai}, J.-M., {Liu}, H.-T., {Chen}, L., \& {Liao}, N.-H. 2012,
  Research in Astronomy and Astrophysics, 12, 1475

\bibitem[{{Fiorucci} {et~al.}(2004){Fiorucci}, {Ciprini}, \&
  {Tosti}}]{2004A&A...419...25F}
{Fiorucci}, M., {Ciprini}, S., \& {Tosti}, G. 2004, \aap, 419, 25

\bibitem[{{Flesch} \& {Hardcastle}(2004)}]{2004A&A...427..387F}
{Flesch}, E. \& {Hardcastle}, M.~J. 2004, \aap, 427, 387

\bibitem[{{Fossati} {et~al.}(1998){Fossati}, {Maraschi}, {Celotti}, {Comastri},
  \& {Ghisellini}}]{1998MNRAS.299..433F}
{Fossati}, G., {Maraschi}, L., {Celotti}, A., {Comastri}, A., \& {Ghisellini},
  G. 1998, \mnras, 299, 433

\bibitem[{{Ghisellini} {et~al.}(2002){Ghisellini}, {Celotti}, \&
  {Costamante}}]{2002A&A...386..833G}
{Ghisellini}, G., {Celotti}, A., \& {Costamante}, L. 2002, \aap, 386, 833

\bibitem[{{Ghisellini} {et~al.}(1998){Ghisellini}, {Celotti}, {Fossati},
  {Maraschi}, \& {Comastri}}]{1998MNRAS.301..451G}
{Ghisellini}, G., {Celotti}, A., {Fossati}, G., {Maraschi}, L., \& {Comastri},
  A. 1998, \mnras, 301, 451

\bibitem[{{Giommi} \& {Padovani}(1994)}]{1994MNRAS.268L..51G}
{Giommi}, P. \& {Padovani}, P. 1994, \mnras, 268, L51

\bibitem[{{Gregory} \& {Condon}(1991)}]{1991ApJS...75.1011G}
{Gregory}, P.~C. \& {Condon}, J.~J. 1991, \apjs, 75, 1011

\bibitem[{{Gregory} {et~al.}(1996){Gregory}, {Scott}, {Douglas}, \&
  {Condon}}]{1996ApJS..103..427G}
{Gregory}, P.~C., {Scott}, W.~K., {Douglas}, K., \& {Condon}, J.~J. 1996,
  \apjs, 103, 427

\bibitem[{{Gregory} {et~al.}(1994){Gregory}, {Vavasour}, {Scott}, \&
  {Condon}}]{1994ApJS...90..173G}
{Gregory}, P.~C., {Vavasour}, J.~D., {Scott}, W.~K., \& {Condon}, J.~J. 1994,
  \apjs, 90, 173

\bibitem[{{Griffith} {et~al.}(1995){Griffith}, {Wright}, {Burke}, \&
  {Ekers}}]{1995ApJS...97..347G}
{Griffith}, M.~R., {Wright}, A.~E., {Burke}, B.~F., \& {Ekers}, R.~D. 1995,
  \apjs, 97, 347

\bibitem[{{Heidt} \& {Nilsson}(2011)}]{2011A&A...529A.162H}
{Heidt}, J. \& {Nilsson}, K. 2011, \aap, 529, A162

\bibitem[{{H.E.S.S.~Collaboration} {et~al.}(2013){H.E.S.S.~Collaboration},
  {Abramowski}, {Acero}, {Aharonian}, {Akhperjanian}, {Ang{\"u}ner}, {Anton},
  {Balenderan}, {Balzer}, {Barnacka}, \& et~al.}]{2013A&A...554A..72H}
{H.E.S.S.~Collaboration}, {Abramowski}, A., {Acero}, F., {et~al.} 2013, \aap,
  554, A72

\bibitem[{{Hovatta} {et~al.}(2014){Hovatta}, {Pavlidou}, {King}, {Mahabal},
  {Sesar}, {Dancikova}, {Djorgovski}, {Drake}, {Laher}, {Levitan},
  {Max-Moerbeck}, {Ofek}, {Pearson}, {Prince}, {Readhead}, {Richards}, \&
  {Surace}}]{2014MNRAS.439..690H}
{Hovatta}, T., {Pavlidou}, V., {King}, O.~G., {et~al.} 2014, \mnras, 439, 690

\bibitem[{{Isobe} {et~al.}(1990){Isobe}, {Feigelson}, {Akritas}, \&
  {Babu}}]{1990ApJ...364..104I}
{Isobe}, T., {Feigelson}, E.~D., {Akritas}, M.~G., \& {Babu}, G.~J. 1990, \apj,
  364, 104

\bibitem[{{Jones} {et~al.}(2009){Jones}, {Read}, {Saunders}, {Colless},
  {Jarrett}, {Parker}, {Fairall}, {Mauch}, {Sadler}, {Watson}, {Burton},
  {Campbell}, {Cass}, {Croom}, {Dawe}, {Fiegert}, {Frankcombe}, {Hartley},
  {Huchra}, {James}, {Kirby}, {Lahav}, {Lucey}, {Mamon}, {Moore}, {Peterson},
  {Prior}, {Proust}, {Russell}, {Safouris}, {Wakamatsu}, {Westra}, \&
  {Williams}}]{2009MNRAS.399..683J}
{Jones}, D.~H., {Read}, M.~A., {Saunders}, W., {et~al.} 2009, \mnras, 399, 683

\bibitem[{{Kalberla} {et~al.}(2005){Kalberla}, {Burton}, {Hartmann}, {Arnal},
  {Bajaja}, {Morras}, \& {P{\"o}ppel}}]{2005A&A...440..775K}
{Kalberla}, P.~M.~W., {Burton}, W.~B., {Hartmann}, D., {et~al.} 2005, \aap,
  440, 775

\bibitem[{{Kapanadze}(2013)}]{2013AJ....145...31K}
{Kapanadze}, B.~Z. 2013, \aj, 145, 31

\bibitem[{{Karlen Shahinyan for the VERITAS
  Collaboration}(2015)}]{2015arXiv150203016K}
{Karlen Shahinyan for the VERITAS Collaboration}. 2015, 2014 Fermi Symposium
  proceedings, eConf C14102

\bibitem[{{Kovalev} {et~al.}(2009){Kovalev}, {Aller}, {Aller}, {Homan},
  {Kadler}, {Kellermann}, {Kovalev}, {Lister}, {McCormick}, {Pushkarev}, {Ros},
  \& {Zensus}}]{2009ApJ...696L..17K}
{Kovalev}, Y.~Y., {Aller}, H.~D., {Aller}, M.~F., {et~al.} 2009, \apjl, 696,
  L17

\bibitem[{{Laurent-Muehleisen} {et~al.}(1999){Laurent-Muehleisen}, {Kollgaard},
  {Feigelson}, {Brinkmann}, \& {Siebert}}]{1999ApJ...525..127L}
{Laurent-Muehleisen}, S.~A., {Kollgaard}, R.~I., {Feigelson}, E.~D.,
  {Brinkmann}, W., \& {Siebert}, J. 1999, \apj, 525, 127

\bibitem[{{Lin} \& {Fan}(2016)}]{2016RAA....16..103L}
{Lin}, C. \& {Fan}, J.-H. 2016, Research in Astronomy and Astrophysics, 16, 103

\bibitem[{{Lindfors} {et~al.}(2016){Lindfors}, {Hovatta}, {Nilsson},
  {Reinthal}, {Fallah Ramazani}, {Pavlidou}, {Max-Moerbeck}, {Richards},
  {Berdyugin}, {Takalo}, {Sillanp{\"a}{\"a}}, \&
  {Readhead}}]{2016A&A...593A..98L}
{Lindfors}, E.~J., {Hovatta}, T., {Nilsson}, K., {et~al.} 2016, \aap, 593, A98

\bibitem[{{Maraschi} {et~al.}(1992){Maraschi}, {Ghisellini}, \&
  {Celotti}}]{1992ApJ...397L...5M}
{Maraschi}, L., {Ghisellini}, G., \& {Celotti}, A. 1992, \apjl, 397, L5

\bibitem[{{Massaro} {et~al.}(2015){Massaro}, {Maselli}, {Leto}, {Marchegiani},
  {Perri}, {Giommi}, \& {Piranomonte}}]{2015Ap&SS.357...75M}
{Massaro}, E., {Maselli}, A., {Leto}, C., {et~al.} 2015, \apss, 357, 75

\bibitem[{{Massaro} {et~al.}(2004){Massaro}, {Perri}, {Giommi}, \&
  {Nesci}}]{2004A&A...413..489M}
{Massaro}, E., {Perri}, M., {Giommi}, P., \& {Nesci}, R. 2004, \aap, 413, 489

\bibitem[{{Massaro} \& {D'Abrusco}(2016)}]{2016ApJ...827...67M}
{Massaro}, F. \& {D'Abrusco}, R. 2016, \apj, 827, 67

\bibitem[{{Massaro} {et~al.}(2013{\natexlab{a}}){Massaro}, {D'Abrusco},
  {Paggi}, {Masetti}, {Giroletti}, {Tosti}, {Smith}, \&
  {Funk}}]{2013ApJS..206...13M}
{Massaro}, F., {D'Abrusco}, R., {Paggi}, A., {et~al.} 2013{\natexlab{a}},
  \apjs, 206, 13

\bibitem[{{Massaro} {et~al.}(2011){Massaro}, {Paggi}, {Elvis}, \&
  {Cavaliere}}]{2011ApJ...739...73M}
{Massaro}, F., {Paggi}, A., {Elvis}, M., \& {Cavaliere}, A. 2011, \apj, 739, 73

\bibitem[{{Massaro} {et~al.}(2013{\natexlab{b}}){Massaro}, {Paggi}, {Errando},
  {D'Abrusco}, {Masetti}, {Tosti}, \& {Funk}}]{2013ApJS..207...16M}
{Massaro}, F., {Paggi}, A., {Errando}, M., {et~al.} 2013{\natexlab{b}}, \apjs,
  207, 16

\bibitem[{{Melia} \& {Konigl}(1989)}]{1989ApJ...340..162M}
{Melia}, F. \& {Konigl}, A. 1989, \apj, 340, 162

\bibitem[{{Mirzoyan}(2015{\natexlab{a}})}]{2015ATel.7080....1M}
{Mirzoyan}, R. 2015{\natexlab{a}}, The Astronomer's Telegram, 7080

\bibitem[{{Mirzoyan}(2015{\natexlab{b}})}]{2015ATel.7844....1M}
{Mirzoyan}, R. 2015{\natexlab{b}}, The Astronomer's Telegram, 7844

\bibitem[{{Mirzoyan}(2016{\natexlab{a}})}]{2016ATel.9582....1M}
{Mirzoyan}, R. 2016{\natexlab{a}}, The Astronomer's Telegram, 9582

\bibitem[{{Mirzoyan}(2016{\natexlab{b}})}]{2016ATel.9267....1M}
{Mirzoyan}, R. 2016{\natexlab{b}}, The Astronomer's Telegram, 9267

\bibitem[{{Monet} {et~al.}(2003){Monet}, {Levine}, {Canzian}, {Ables}, {Bird},
  {Dahn}, {Guetter}, {Harris}, {Henden}, {Leggett}, {Levison}, {Luginbuhl},
  {Martini}, {Monet}, {Munn}, {Pier}, {Rhodes}, {Riepe}, {Sell}, {Stone},
  {Vrba}, {Walker}, {Westerhout}, {Brucato}, {Reid}, {Schoening}, {Hartley},
  {Read}, \& {Tritton}}]{2003AJ....125..984M}
{Monet}, D.~G., {Levine}, S.~E., {Canzian}, B., {et~al.} 2003, \aj, 125, 984

\bibitem[{{M{\"u}cke} \& {Protheroe}(2001)}]{2001APh....15..121M}
{M{\"u}cke}, A. \& {Protheroe}, R.~J. 2001, Astroparticle Physics, 15, 121

\bibitem[{{Mukherjee} \& {VERITAS Collaboration}(2017)}]{2017ATel10051....1M}
{Mukherjee}, R. \& {VERITAS Collaboration}. 2017, The Astronomer's Telegram,
  No.~10051, 51

\bibitem[{{Nemmen} {et~al.}(2012){Nemmen}, {Georganopoulos}, {Guiriec},
  {Meyer}, {Gehrels}, \& {Sambruna}}]{2012Sci...338.1445N}
{Nemmen}, R.~S., {Georganopoulos}, M., {Guiriec}, S., {et~al.} 2012, Science,
  338, 1445

\bibitem[{{Nieppola} {et~al.}(2011){Nieppola}, {Tornikoski}, {Valtaoja},
  {Le{\'o}n-Tavares}, {Hovatta}, {L{\"a}hteenm{\"a}ki}, \&
  {Tammi}}]{2011A&A...535A..69N}
{Nieppola}, E., {Tornikoski}, M., {Valtaoja}, E., {et~al.} 2011, \aap, 535, A69

\bibitem[{{Nilsson} {et~al.}(2007){Nilsson}, {Pasanen}, {Takalo}, {Lindfors},
  {Berdyugin}, {Ciprini}, \& {Pforr}}]{2007A&A...475..199N}
{Nilsson}, K., {Pasanen}, M., {Takalo}, L.~O., {et~al.} 2007, \aap, 475, 199

\bibitem[{{Nilsson} {et~al.}(2003){Nilsson}, {Pursimo}, {Heidt}, {Takalo},
  {Sillanp{\"a}{\"a}}, \& {Brinkmann}}]{2003A&A...400...95N}
{Nilsson}, K., {Pursimo}, T., {Heidt}, J., {et~al.} 2003, \aap, 400, 95

\bibitem[{{Nilsson} {et~al.}(1999){Nilsson}, {Pursimo}, {Takalo},
  {Sillanp{\"a}{\"a}}, {Pietil{\"a}}, \& {Heidt}}]{1999PASP..111.1223N}
{Nilsson}, K., {Pursimo}, T., {Takalo}, L.~O., {et~al.} 1999, \pasp, 111, 1223

\bibitem[{{Padovani} \& {Giommi}(2015)}]{2015MNRAS.446L..41P}
{Padovani}, P. \& {Giommi}, P. 2015, \mnras, 446, L41

\bibitem[{{Padovani} \& {Resconi}(2014)}]{2014MNRAS.443..474P}
{Padovani}, P. \& {Resconi}, E. 2014, \mnras, 443, 474

\bibitem[{{Pita} {et~al.}(2014){Pita}, {Goldoni}, {Boisson}, {Lenain}, {Punch},
  {G{\'e}rard}, {Hammer}, {Kaper}, \& {Sol}}]{2014A&A...565A..12P}
{Pita}, S., {Goldoni}, P., {Boisson}, C., {et~al.} 2014, \aap, 565, A12

\bibitem[{{Planck Collaboration} {et~al.}(2014){Planck Collaboration}, {Ade},
  {Aghanim}, {Armitage-Caplan}, {Arnaud}, {Ashdown}, {Atrio-Barandela},
  {Aumont}, {Baccigalupi}, {Banday}, \& et~al.}]{2014A&A...571A..16P}
{Planck Collaboration}, {Ade}, P.~A.~R., {Aghanim}, N., {et~al.} 2014, \aap,
  571, A16

\bibitem[{{Sbarufatti} {et~al.}(2005){Sbarufatti}, {Treves}, \&
  {Falomo}}]{2005ApJ...635..173S}
{Sbarufatti}, B., {Treves}, A., \& {Falomo}, R. 2005, \apj, 635, 173

\bibitem[{{Schlafly} \& {Finkbeiner}(2011)}]{2011ApJ...737..103S}
{Schlafly}, E.~F. \& {Finkbeiner}, D.~P. 2011, \apj, 737, 103

\bibitem[{{Schulz} {et~al.}(2015){Schulz}, {Kadler}, {Ros}, {Eisenacher
  Glawion}, {Bach}, {Els{\"a}sser}, {Grossberger}, {Kreykenbohm}, {Mannheim},
  {M{\"u}ller}, {Tr{\"u}stedt}, \& {Wilms}}]{2015arXiv150203559S}
{Schulz}, R., {Kadler}, M., {Ros}, E., {et~al.} 2015, ArXiv e-prints:
  1502.03559, Proceedings of the 12th European VLBI Network Symposium and Users
  Meeting - EVN 2014, 7-10 October 2014, Cagliari, Italy. Published online in
  PoS, ID.109

\bibitem[{{Sikora} {et~al.}(1994){Sikora}, {Begelman}, \&
  {Rees}}]{1994ApJ...421..153S}
{Sikora}, M., {Begelman}, M.~C., \& {Rees}, M.~J. 1994, \apj, 421, 153

\bibitem[{{Voges} {et~al.}(1999){Voges}, {Aschenbach}, {Boller},
  {Br{\"a}uninger}, {Briel}, {Burkert}, {Dennerl}, {Englhauser}, {Gruber},
  {Haberl}, {Hartner}, {Hasinger}, {K{\"u}rster}, {Pfeffermann}, {Pietsch},
  {Predehl}, {Rosso}, {Schmitt}, {Tr{\"u}mper}, \&
  {Zimmermann}}]{1999A&A...349..389V}
{Voges}, W., {Aschenbach}, B., {Boller}, T., {et~al.} 1999, \aap, 349, 389

\bibitem[{{Wagner}(2008)}]{2008MNRAS.385..119W}
{Wagner}, R.~M. 2008, \mnras, 385, 119

\bibitem[{{Wright}(2006)}]{2006PASP..118.1711W}
{Wright}, E.~L. 2006, \pasp, 118, 1711

\bibitem[{{Wright} {et~al.}(2010){Wright}, {Eisenhardt}, {Mainzer}, {Ressler},
  {Cutri}, {Jarrett}, {Kirkpatrick}, {Padgett}, {McMillan}, {Skrutskie},
  {Stanford}, {Cohen}, {Walker}, {Mather}, {Leisawitz}, {Gautier}, {McLean},
  {Benford}, {Lonsdale}, {Blain}, {Mendez}, {Irace}, {Duval}, {Liu}, {Royer},
  {Heinrichsen}, {Howard}, {Shannon}, {Kendall}, {Walsh}, {Larsen}, {Cardon},
  {Schick}, {Schwalm}, {Abid}, {Fabinsky}, {Naes}, \&
  {Tsai}}]{2010AJ....140.1868W}
{Wright}, E.~L., {Eisenhardt}, P.~R.~M., {Mainzer}, A.~K., {et~al.} 2010, \aj,
  140, 1868

\end{thebibliography}

\clearpage
\begin{sidewaystable}
\caption{First 5 rows of the TeV BL Lacs properties online table.}              
\label{tab1}      
\centering                                    
\begin{tabular}{lcccccccccccc}          
\hline\hline                       
\multirow{2}{*}{Source name}			&\multirow{2}{*}{z}	 &$L_R$	&$L_I$	&$L_O$	&$L_X$	&$L_\gamma$		&$L_{VHE,H}$ &$L_{VHE,L}$ &$L_{VHE,SD}$&\multirow{2}{*}{Ref.}\\ 
\cline{3-10}
	&			 &\multicolumn{8}{c}{[erg/s]}&\\ 
\hline  
BZB~J0013-1854 &0.095			&\num{4.78E+40} &\num{2.02E+43} &\num{9.15E+42} &\num{1.58E+44} &\num{4.64E+44}& & &\num{2.12E+43}&1,5,6,8,13,14\\
KUV~00311-1938 &0.470\tablefootmark{a}  &\num{5.27E+41} &\num{6.20E+44} &\num{7.57E+45} &\num{5.25E+45} &\num{1.76E+47}& & &\num{8.61E+44}&2,3,6,9,13,15\\
1ES~0033+595   &0.240\tablefootmark{b}	&\num{1.08E+42} &\num{1.12E+45} &\num{1.16E+45} &\num{1.11E+46} &\num{3.89E+46}& & &\num{2.79E+44}&4,3,7,10,13,16\\
RGB~J0152+017  &0.080		 	&\num{4.67E+40} &\num{1.98E+43} &\num{5.53E+43} &\num{4.87E+43} &\num{1.03E+45}& & &\num{3.50E+43}&3,3,6,11,13,11\\
3C~66A	       &0.335\tablefootmark{c}	&\num{3.34E+43} &\num{8.03E+45} &\num{1.83E+46} &\num{2.02E+45} &\num{6.06E+47}&\num{5.14E+45}&\num{9.15E+44}&&4,3,7,12,13,17\\

\hline
\end{tabular}
\tablefoot{
\tablefoottext{a}{Lower limit based on spectroscopy.}
\tablefoottext{b}{Lower limit based on non-detection of host galaxy.}
\tablefoottext{c}{Lower limit based on detection of $Ly\alpha$ forests.}}
\tablebib{
(1) \citet{2009MNRAS.399..683J}; 
(2) \citet{2014A&A...565A..12P}; 
(3) \citet{2011ApJ...743..171A}; 
(4) \citet{2016A&A...593A..98L}; 
(5) \citet{1995ApJS...97..347G}; 
(6) \citet{2004A&A...427..387F}; 
(7) Nilsson et al. (in prep.); 
(8) \citet{2011ApJ...739...73M}; 
(9) This work; 
(10) \citet{2005A&A...433.1163D}; 
(11) \citet{2008A&A...481L.103A}; 
(12) \citet{2010ApJ...716...30A}; 
(13) \citet{2015ApJS..218...23A}; 
(14) \citet{2013A&A...554A..72H}; 
(15) \citet{10.1063/1.4772304}; 
(16) \citet{2015MNRAS.446..217A}; 
(17) \citet{2011ApJ...726...58A}
}
\end{sidewaystable}

\begin{sidewaystable}
\caption{Promising TeV candidates and the example of online version of non-TeV BL Lacs properties.}              
\label{tab2}      
\centering                                    
\begin{tabular}{lcccccccccc}          
\hline\hline 
\multirow{2}{*}{Source name}	&\multirow{2}{*}{z} &$S_R$ &$S_I$ &$S_O$ &$S_X$ &$S_\gamma$ &$S_{VHE,min}$&$S_{VHE,med}$&$S_{VHE,max}$&\multirow{2}{*}{Notes\tablefootmark{a}}	\\   
\cline{3-10}
	&				&\multicolumn{7}{c}{$\rm [erg/cm^2/s]$}&&\\
\hline  
3C~371                     & 0.046 & \num{9.15E-14} & \num{1.06E-11} & \num{2.26E-11} & \num{2.50E-12} & \num{4.06E-10} & \num{3.16E-12} & \num{8.79E-12} & \num{2.60E-11} &$c_3,c_2$ \\ 
PKS~0829+046               & 0.174 & \num{6.02E-14} & \num{8.05E-12} & \num{9.96E-12} & \num{3.00E-12} & \num{3.95E-10} & \num{1.50E-12} & \num{3.21E-12} & \num{8.40E-12} &$c_2,c_6$ \\ 
1RXS~J195815.6-301119      & 0.119 & \num{6.21E-15} & \num{1.80E-12} & \num{4.64E-11} & \num{6.66E-12} & \num{1.72E-10} & \num{1.14E-12} & \num{3.95E-12} & \num{1.99E-11} &$c_6,c_8$ \\ 
B2~1811+31                 & 0.117 & \num{9.26E-15} & \num{4.41E-12} & \num{5.89E-12} & \num{1.23E-13} & \num{2.68E-10} & \num{1.07E-12} & \num{2.56E-12} & \num{8.27E-12} &$c_1,c_2$ \\ 
1H~1914-194                & 0.137 & \num{2.33E-14} & \num{4.74E-12} & \num{8.78E-12} & \num{7.90E-13} & \num{1.61E-10} & \num{1.05E-12} & \num{2.26E-12} & \num{5.42E-12} &$c_3,c_2$ \\ 
PMN~J0444-6014             & 0.097 & \num{1.79E-15} & \num{9.39E-13} & \num{4.94E-11} &       -        & \num{2.80E-10} & \num{9.31E-13} & \num{5.42E-12} & \num{2.41E-11} &$c_4,c_6$ \\ 
B2~0806+35                 & 0.083 & \num{7.61E-15} & \num{1.01E-11} & \num{3.23E-12} & \num{2.03E-12} & \num{1.12E-10} & \num{6.70E-13} & \num{2.40E-12} & \num{6.03E-12} &$c_1,c_6$ \\ 
4C~+42.22                  & 0.059 & \num{2.33E-14} & \num{3.43E-12} & \num{3.31E-12} & \num{2.09E-12} & \num{8.01E-11} & \num{6.64E-13} & \num{2.44E-12} & \num{8.78E-12} &$c_3,c_2$ \\ 
TXS~0210+515               & 0.049 & \num{1.41E-14} & \num{6.68E-12} & \num{1.68E-11} & \num{1.46E-11} & \num{7.42E-11} & \num{6.52E-13} & \num{5.64E-12} & \num{1.78E-11} &$c_3,c_8$ \\ 
\hline
\end{tabular}
\tablefoot{
\tablefoottext{a}{The prediction function(s), listed in table \ref{tab4}, which are used to calculate $S_{VHE,med}$.}}
\end{sidewaystable}

\end{document}